\documentclass[a4,11pt]{article}
\usepackage[dvipdfmx]{graphicx}
\usepackage{amsmath,amssymb,amsfonts,color,graphics,amscd,amsfonts}
\newcommand{\beq}{\begin{eqnarray}}
\newcommand{\eeq}{\end{eqnarray}}

\newcommand{\tr}{\mathop{\mathrm{tr}}}

\allowdisplaybreaks[4]

\date{}
\begin{document}

\begin{flushright} 
\today\\
YITP-14-71

SU-ITP-14/22

\end{flushright} 

\vspace{0.1cm}

\begin{center}
  {\LARGE

Taming the pion condensation in  QCD at finite baryon density

 
}
\end{center}
\vspace{0.1cm}
\vspace{0.1cm}
\begin{center}
Shinya A{\sc oki}$^{a}$\footnote{saoki@yukawa.kyoto-u.ac.jp},   
 Masanori H{\sc anada}$^{b,a,c}$\footnote{hanada@yukawa.kyoto-u.ac.jp}
and 
Atsushi N{\sc akamura}$^{d}$\footnote{nakamura@riise.hiroshima-u.ac.jp}

\vspace{0.5cm}

$^a${\it Yukawa Institute for Theoretical Physics, Kyoto University,\\
Kitashirakawa Oiwakecho, Sakyo-ku, Kyoto 606-8502, Japan}	

$^b${\it Stanford Institute for Theoretical Physics,
Stanford University, Stanford, CA 94305, USA}

$^c${\it The Hakubi Center for Advanced Research, Kyoto University,\\
Yoshida Ushinomiyacho, Sakyo-ku, Kyoto 606-8501, Japan}

$^d${\it Research Institute for Information Science and Education,
Hiroshima University,\\ 
Higashi-Hiroshima 739-8527 Japan}

\end{center}

\vspace{1.5cm}

\begin{center}
  {\bf Abstract}
\end{center}
In the Monte Carlo study of QCD at finite baryon density based upon the phase reweighting method, the pion condensation in the phase-quenched theory and associated zero-mode prevent us to go to the low-temperature high-density region. 
We propose a  method to circumvent them by a simple modification of the density of state method. 
We first argue that the standard version of the density of state method, which is invented  to solve the overlapping problem, 
is effective only for a certain `good' class of observables. 
We then  modify it   so as to solve  the overlap problem for `bad' observables as well. 
While, in the standard version of the density of state method, we  usually constrain
an observable we are interested in,   
we fix a different  observable in our new method which has a sharp peak at some particular value characterizing  the correct vacuum of the target theory.
In the finite-density QCD, such an observable is the pion condensate. 
The average phase becomes vanishingly small as the value of the pion condensate becomes large, 
hence it is enough to consider configurations with $\pi^+\simeq 0$, where the zero mode does not appear. 
We demonstrate an effectiveness of our method by using a toy model (the chiral random matrix theory) which captures the properties of finite-density QCD qualitatively. 
We also argue how to apply our method to other theories including finite-density QCD. 
Although the example we study numerically is based on the phase reweighting method, the same idea can be applied to more general reweighting methods and
we show how this idea can be applied to find a possible QCD critical point.

\newpage

\section{Introduction: importance sampling and {\it un}-importance sampling}
\hspace{0.51cm}
The sign problem is a severe obstacle for Monte Carlo methods based on the importance sampling, and
it prevents us, for example, from studying lattice QCD at finite baryon density directly by Monte Carlo simulations,
since the fermion determinant becomes complex at a finite baryon chemical potential. 
(For an introductory review from lattice perspective, see \cite{Muroya:2003qs}. 
A review from the point of view of nuclear theory can be found in \cite{Fukushima:2010bq}.) 
Several methods have been proposed to overcome this difficulty (for various previous attempts, see e.g.  
\cite{Nakamura:1984uz,Barbour:1988ax,Fodor:2001pe,Parisi:1984cs,Cristoforetti:2012su,de Forcrand:2002ci,Allton:2002zi}), 
and some of them are based on the phase-reweighting technique, which, however, fail to work at high density due to the unphysical pion condensation. 
 In this paper we propose a new method to tame the pion condensation problem of  reweighting methods,
 whose basic idea  can also be applied to some classes of sign problems. 
As a bonus, a zero mode associated with the unphysical pion condensation is eliminated.

Let us begin with identifying the physical origin of the sign problem. 
We consider a field theory on Euclidean spacetime with a complex action, 
\begin{eqnarray}
S
=
S_R+iS_I. 
\end{eqnarray}
Then the path-integral weight ${\rm e}^{-S}$ is not real and positive anymore, and hence the importance sampling cannot be applied as it is. 
Therefore one performs the importance sampling by using a real and positive weight which `approximates' the complex weight 
and take into account the effect of the non-positivity by using so-called reweighting methods. 
The simplest example is the phase-reweighting method, in which the phase-quenched weight ${\rm e}^{-S_R}$ is adopted; 
the expectation value of an operator $\hat{O}$ in the full theory is obtained by using an identity 
\begin{eqnarray}
\langle \hat{O}\rangle_{\rm full}
=
\frac{\langle {\rm e}^{iS_I}\cdot\hat{O}\rangle_{\rm P.Q.}}{\langle {\rm e}^{iS_I}\rangle_{\rm P.Q.}},  
\end{eqnarray}
where $\langle\ \cdot\ \rangle_{\rm full}$ and $\langle\ \cdot\ \rangle_{\rm P.Q.}$ stand for expectation values 
in the full and the phase-quenched theories, respectively\footnote{
\begin{eqnarray}
\langle \hat{O}\rangle_{\rm full}
\equiv
\frac{\int [d\phi]\hat{O}[\phi]{\rm e}^{-S[\phi]}}{\int [d\phi]{\rm e}^{-S[\phi]}}
\end{eqnarray}
and 
\begin{eqnarray}
\langle \hat{O}\rangle_{\rm P.Q.}
\equiv
\frac{\int [d\phi]\hat{O}[\phi]{\rm e}^{-S_R[\phi]}}{\int [d\phi]{\rm e}^{-S_R[\phi]}}. 
\end{eqnarray}
}. 
Then the right hand side is calculable in principle. 
In practice, however, both $\langle {\rm e}^{iS_I}\rangle_{\rm P.Q.}$ and $\langle {\rm e}^{iS_I}\cdot\hat{O}\rangle_{\rm P.Q.}$ can become extremely small in some cases and 
then the right hand side is essentially $0/0$, which is not easy to  evaluate numerically. 
This is the sign problem. 
 
The sign problem becomes even severer when the vacua of the full and phase-quenched theories are different; 
this is so-called `overlap problem'. 
In order to understand it, let us consider a certain observable $\hat{O}$ which characterizes the vacua of these two theories;  
the vacua are characterized by $\langle \hat{O}\rangle_{\rm full}=K_{\rm full}$ and $\langle \hat{O}\rangle_{\rm P.Q.}=K_{\rm P.Q.}$, where 
$K_{\rm full}\neq K_{\rm P.Q.}$ in general. 
Let us denote the histogram of $\hat{O}$ in the phase-quenched theory as $\rho_{\rm P.Q.}(x)$. 
It peaks around $x=K_{\rm P.Q.}$. The `histogram' in the full theory is proportional to $\rho_{\rm P.Q.}(x)\cdot\langle {\rm e}^{iS_I}\rangle_{x}$, 
where  $\langle {\rm e}^{iS_I}\rangle_{x}$ is the average phase factor with the value of $\hat{O}$ fixed to $x$. 
Since $\rho_{\rm full}(x)\sim\rho_{\rm P.Q.}(x)\cdot\langle {\rm e}^{iS_I}\rangle_{x}$ peaks around $K_{\rm full}$, 
the phase factor $\langle {\rm e}^{iS_I}\rangle_{x}\sim \rho_{\rm full}(x)/\rho_{\rm P.Q.}(x)$ is vanishingly small around $x=K_{\rm P.Q.}\neq K_{\rm full}$.  
(This point is clearly demonstrated in \cite{Splittorff:2006fu} by using a solvable model.) 
This means that, although most configurations sampled in the phase quenched simulation are around $x=K_{\rm P.Q.}$, 
their contribution vanishes due to huge sign fluctuation, and the true peak of the full theory appears from the tail of  $\rho_{\rm P.Q.}(x)$. 
In other words the phase-quenched simulation is the {\it un}-importance sampling, in the sense that the most of computational resources
are wasted to sample un-important  configurations. 
In fact it is even worse -- 
the sign fluctuation becomes violent in order to erase un-important  configurations,  and 
unless one has huge amount of  configurations so that vanishingly small value of the phase factor can be measured precisely, 
the error bar becomes large; essentially the only contribution of the un-important samples is to make the error bar larger. 
Such a waste of computational resources, which arises  due to the lack of the overlap  between vacua in full and phase-quenched theories, 
is the overlap problem. 

In terms of the above general argument, 
we consider the massless two-flavor QCD with the finite baryon chemical potential (QCD$_B$).
In this theory, two quarks (up and down) has the same value of the chemical potential $\mu$, which
coupled to the baryon number of quarks, $+1/3$.
The partition function in Euclidean space-time is given by 
\begin{eqnarray}
Z_{\rm full}
=
\int [dA_\mu] \left[\det(\gamma^\mu D_\mu(A)+\mu\gamma^4)\right]^2 {\rm e}^{-S_G(A)}
\end{eqnarray}
where $D_\mu$ is the gauge covariant derivative acting on quark fields $\psi=u,d$, $D_\mu(A)\psi=(\partial_\mu-iA_\mu)\psi$ with 
the gauge field $A_\mu$, and $S_G(A)$ is the action for the gauge field. 
The determinant factor satisfies 
\begin{eqnarray}
\left[\det(\gamma^\mu D_\mu(A)+\mu\gamma^4)\right]^\ast
=
\det(\gamma^\mu D_\mu(A)-\mu\gamma^4), 
\end{eqnarray}
and hence it is complex at $\mu\not= 0$, so that the sign problem exists in QCD$_B$. 
The phase quenched theory is described by the partition function,
\begin{eqnarray}
Z_{\rm P.Q.}
&=&
\int [dA_\mu] \left|\det(\gamma^\mu D_\mu(A)+\mu\gamma^4)\right|^2 {\rm e}^{-S_G(A)}\nonumber \\
&=&
\int [dA_\mu] \det(\gamma^\mu D_\mu(A)+\mu\gamma^4)\cdot\det(\gamma^\mu D_\mu(A)-\mu\gamma^4) {\rm e}^{-S_G(A)}. 
\label{eq:ZPQ}
\end{eqnarray}
This theory is QCD with a finite isospin chemical potential (QCD$_I$), in which up and down quarks have chemical potential $+\mu$ and $-\mu$, respectively.  
Hence this chemical potential couples to the isospin number,  $+1/2$ for up and $-1/2$ for down.  
In the full theory, nothing happens until the nucleon, whose mass is about 1 GeV, condenses. 
On the other hand, in the phase quenched theory, the massless charged pion $\pi^+=\bar{d}\gamma^5 u$ 
condenses as soon as $\mu$ is turned on. 
Therefore the overlapping problem arises due to the pion condensation.  

In this paper we propose a simple way to tame the sign problem caused by  the overlap problems associated with the pion condensation in the phase quenched theory.
We first notice that,  if one eliminates the pion condensate by hand (for example by adding delta-function like potential), two theories, QCD$_B$ and QCD$_I$, 
become equivalent when the number of colors $N_c$ is sent to infinity \cite{Cherman:2010jj,Hanada:2011ju}  ($N_c=3$ is the usual QCD), which
means that the overlap problem is just a $1/N_c$ effect if we fix the pion condensate
\footnote{
The equivalence at $N_c=\infty$ holds if one takes the massless limit after taking the large-$N_c$. 
Strictly speaking, at very large $\mu$, other isospin-charged particles like the rho-meson would condense and lead to the overlap problem, 
and then  their condensates must be fixed to be zero. 
}${}^,$\footnote{The remaining overlap problem is due to the gas of pions. 
The overlap problem is mild as long as the pion does not condensate, and even the phase quench is exact at large-$N_c$. 
We will comment on this point later.}. 
This consideration leads to our main idea that the overlap problem can be avoided by pinning down an appropriate observable, 
which characterizes the difference between full and phase quenched theories (in  the case of the finite density QCD, the pion condensate), to the right value 
(zero pion condensation in QCD$_B$), and the sign fluctuation becomes milder there. 
Away from the correct vacuum, the sign fluctuation becomes severer. This is not drawback anymore, 
because the severe sign fluctuation is simply telling us that such configurations are not important. 
When the sign fluctuation becomes severer, we do not have to measure the average sign. 
Rather, we can safely omit such configurations. The sign fluctuation is not a problem anymore, rather it reduces numerical costs of our simulations. 
Furthermore, this methods automatically avoids a zero mode associated with the unphysical pion condensation, 
since we do not have to consider the large-$\pi^+$ region, where the zero mode appears.

Our method is a natural generalization of the density of state method. 
In order to illustrate the advantage of our method, we first review the traditional density of state method, 
explain what is good and what is insufficient, and then introduce our method. 
(Our method could be regarded as a simplified version of the multi-parameter factorization method \cite{Anagnostopoulos:2010ux}, 
which has been applied for a supersymmetric matrix model\footnote{
We would like to thank J.~Nishimura for a comment on this point. 
}. 
We also explain how our method can be combined with the multi-parameter factorization.)

In this paper, we demonstrate our idea taking the chiral random matrix theory (RMT) as  a simpler example. 
Because RMT is analytically solvable and computationally much cheaper than QCD,  
we can test the method thoroughly. 
We explain basic ideas in Sec.~\ref{sec:method} using the chiral RMT.
Note that our main idea does not rely on the detail of the theory  
and the method can be generalized to QCD and other theories. 
In Sec.~\ref{sec:simulation} we give simulation results of the chiral RMT to show how our method works. 
In Sec.~\ref{sec:QCD} we briefly discuss strategies for the finite-density  QCD  using our method, and give
more generic reweighing method in Sec. \ref{sec:general}.
Our conclusion and discussion are given in Sec.\ref{sec:conclusion}.

\section{Methodology}\label{sec:method}
\hspace{0.51cm}
In this section we explain our method  using the chiral RMT as a concrete example. 
\subsection{$\beta=2$ RMT}
\hspace{0.51cm}
The action of the $\beta=2$ RMT \cite{Shuryak:1992pi,Verbaarschot:1994qf} with chemical potential \cite{Stephanov:1996ki} is given by 
\begin{eqnarray}
Z=\int d\Phi d\Psi \ {\rm e}^{-S}, \qquad S=S_{B}+S_{F},
\end{eqnarray}
where 
\begin{eqnarray}
S_B
=N \tr \Phi\Phi^\dagger, \qquad
S_F
=
\sum_{f=1}^{N_f}
\bar{\Psi}_f {\cal D}_f\Psi_f, 
\end{eqnarray}
and  
\begin{eqnarray}
{\cal D}_f
=
\left(
\begin{array}{cc}
m_f \textbf{1}_{N}& \Phi+\mu_f\textbf{1}_{N} \\
-\Phi^\dagger +\mu_f\textbf{1}_{N} & m_f \textbf{1}_{N}
\end{array}
\right).   
\end{eqnarray}
Here $\Phi$ is $N\times N$ complex matrix. 
From now on we take the number of flavors $N_f$ to be two (up and down quarks). 
We assign $\mu_1=\mu_2=\mu$ for the full theory (finite baryon chemical potential) 
and $\mu_1=+\mu$, $\mu_2=-\mu$ for the phase-quenched theory (isospin chemical potential). 
We call these matrix models as RMT$_B$ and RMT$_I$, respectively. 
Hereafter we will take a massless limit ($m_u=m_d=0$). Therefore  `chiral condensate'' $\langle \bar u u + \bar d d \rangle$ will not be discussed in this paper.

The pion condensate is identically zero unless we introduce a source term. 
We introduce a source term to RMT$_I$ as 
\begin{eqnarray}
\tilde{{\cal D}}
=
\left(
\begin{array}{cc|cc}
0  & \Phi+\mu\textbf{1}_{N} & c\textbf{1}_{N} & 0 \\
-\Phi^\dagger +\mu\textbf{1}_{N} & 0  & 0 & -c\textbf{1}_{N}\\
\hline
-c\textbf{1}_{N}  & 0 & 0 & \Phi-\mu\textbf{1}_{N} \\
0 & c\textbf{1}_{N} & -\Phi^\dagger -\mu\textbf{1}_{N} & 0  \\
\end{array}
\right)
\equiv \left(
\begin{array}{cc}
{\cal D}(\mu) & c\gamma_5 \\
-c\gamma_5 & {\cal D}(-\mu)
\end{array}
\right), 
\end{eqnarray} 
where $c$ is a real number, and
\begin{eqnarray}
\gamma_5
=
\left(
\begin{array}{cc}
\textbf{1}_{N} & 0 \\
0 & -\textbf{1}_{N}
\end{array}
\right).     
\end{eqnarray} 
Then the `pion condensate' is real and satisfy 
\begin{eqnarray}
\pi^+
\equiv
{\rm Tr}[\gamma_5\cdot (\tilde{{\cal D}}^{-1})_{21}]/N 
=
-{\rm Tr}[\gamma_5\cdot (\tilde{{\cal D}}^{-1})_{12}]/N = -\pi^{-}
\end{eqnarray}


As an observable we will measure the baryon density $\nu_B$, which is defined by 
$\langle \nu_B\rangle_B=\langle \bar{u}\gamma^4 u\rangle_B +\langle \bar{d}\gamma^4 d\rangle_B = 2\langle\bar{u}\gamma^4 u \rangle_B=2\langle {\rm Tr}(\gamma^4 {\cal D}^{-1}(+\mu))\rangle_B$, 
where 
\begin{eqnarray}
\gamma^4
=
\left(
\begin{array}{cc}
0 & \textbf{1}_{N} \\
\textbf{1}_{N} & 0
\end{array}
\right).     
\end{eqnarray} 

In QCD, we have
\begin{eqnarray}
\bar\psi^c  D(A^c,\,\mu) \psi^c &=& \bar\psi D(A, -\mu) \psi, \quad
D(A,\mu) = \gamma^\mu D_\mu(A_\mu) + \mu\gamma^4,
\end{eqnarray}
where
the charge conjugations are defined by
\begin{eqnarray}
\psi^c =  C \bar\psi^T  , \quad
\bar\psi^c &=& - \psi^T C^{-1}, \quad
A_\mu^c = - A_\mu^T,
\end{eqnarray}
$C$ is the charge conjugation matrix satisfying $C^{-1} \gamma^\mu C = - (\gamma^\mu)^T$, and $T$ stands for the transpose. This also implies
\begin{equation}
C^{-1}  D(A^c,\mu) C = D(A,-\mu)
\end{equation}
Using these we see that
\begin{eqnarray}
\langle  \bar d\gamma^4 d \rangle_{\rm P.Q.} &=&  
Z^{-1}\int [dA_\mu] \det [ D(A,\mu) D(A,-\mu)] \ \tr [\gamma^4 D^{-1}(A,\mu)]\nonumber \\
&=& -Z^{-1}\int [A_\mu^c] \det [ D(A^c, -\mu) D(A^c,\mu)]\  \tr [\gamma^4 D^{-1}(A^c,-\mu)] = - \langle  \bar d\gamma^4 d \rangle_I .
\end{eqnarray}
Therefore
\begin{eqnarray}
\langle  \nu_B \rangle_{\rm P.Q.} &=&  \langle  \bar u\gamma^4 u \rangle_{\rm P.Q.} 
+\langle  \bar d\gamma^4 d \rangle_{\rm P.Q.} =  \langle  \bar u\gamma^4 u \rangle_I 
-\langle  \bar d\gamma^4 d \rangle_I = \langle  \nu_I \rangle_I ,
\end{eqnarray}
which means  $\nu_B$ in the phase quenched theory can be regarded as the isospin density $\nu_I$ in QCD$_I$.  
It is easy to see explicitly that these properties also hold in the RMT.
\subsection{Standard density of state method: when it works and when it fails}
\hspace{0.51cm}
First let us explain the standard density of state method (in the context of the finite-density QCD, 
see e.g. \cite{Ambjorn:2002pz,Ejiri:2007ga,Ejiri:2008xt}), in order to illustrate  
the essence of our method explained in the next subsection.

Suppose we want to measure a certain quantity $\hat{O}$, for example $\hat O = \nu_B$.
In the density of state method, one first classifies configurations obtained from the phase-quenched simulation 
in terms of  values of $\hat{O}$. Let the number of configurations (or equivalently, the height of the histogram) 
at $x_i<\hat{O}<x_{i+1}$ be $\rho^{(\hat{O})}_i$, and the average sign be $\langle {\rm e}^{iS_I}\rangle^{(\hat{O})}_i$.  
(Here we have assumed the $\hat{O}$ takes only real values for simplicity.) 
Then we have a trivial relation, 
\begin{eqnarray}
\langle {\rm e}^{iS_I}\rangle_{\rm P.Q.}
=
\frac{\sum_i \langle {\rm e}^{iS_I}\rangle^{(\hat{O})}_i\cdot\rho^{(\hat{O})}_i}{\sum_i\rho^{(\hat{O})}_i}.  
\end{eqnarray}
In the same manner, 
\begin{eqnarray}
\langle {\rm e}^{iS_I}\cdot\hat{O}\rangle_{\rm P.Q.}
=
\frac{\sum_i \langle {\rm e}^{iS_I}\cdot\hat{O}\rangle^{(\hat{O})}_i\cdot\rho^{(\hat{O})}_i}{\sum_i\rho^{(\hat{O})}_i}. 
\end{eqnarray}
Therefore the phase reweighting can be done as 
\begin{eqnarray}
\langle \hat{O}\rangle_{\rm full}
=
\frac{\langle {\rm e}^{iS_I}\cdot\hat{O}\rangle_{\rm P.Q.}}{\langle {\rm e}^{iS_I}\rangle_{\rm P.Q.}}
=
\frac{\sum_i \langle {\rm e}^{iS_I}\cdot\hat{O}\rangle^{(\hat{O})}_i\cdot\rho^{(\hat{O})}_i}{\sum_i \langle {\rm e}^{iS_I}\rangle^{(\hat{O})}_i\cdot\rho^{(\hat{O})}_i}.
\label{eq:reweight_naive} 
\end{eqnarray}

In a naive phase-quenched simulation, the configurations are generated with the weight $\rho^{(\hat{O})}_i$, 
which is different from the weight in the full theory $\langle {\rm e}^{iS_I}\rangle^{(\hat{O})}_i\cdot\rho^{(\hat{O})}_i$. 
(This is the {\it overlap problem}.) 
In order to avoid this overlap problem, 
one performs a {\it constrained simulation}\footnote{
In the following sections we explain how to perform a constrained simulation in the case when $\hat{O}$ is the pion condensate. 
One can constrain other quantities in the same manner. 
} at $x_i<\hat{O}<x_{i+1}$ for all $i$'s 
and evaluates \eqref{eq:reweight_naive}. 
This is the density of state method.
Note that 
the sign problem still remains, because one has to measure 
$\langle {\rm e}^{iS_I}\rangle^{(\hat{O})}_i$ and $\langle {\rm e}^{iS_I}\cdot\hat{O}\rangle^{(\hat{O})}_i$. 
This method works when this remaining sign problem is under control. 
For example, if  $\langle {\rm e}^{iS_I}\rangle^{(\hat{O})}_i$ and $\langle {\rm e}^{iS_I}\cdot\hat{O}\rangle^{(\hat{O})}_i$ do not have clear peaks and  are vanishingly small, the remaining sign problem is still serious. 

It is commonly believed that the density of state method solves the overlap problem completely,  
because all the values of $\hat{O}$ are scanned.  In fact this is not really true, because this method is not based on the idea of the importance sampling.  
As we have explained in the introduction, the role of the sign fluctuation is to erase  contributions from the wrong vacuum (the vacuum of the phase quenched theory) 
and to realize  the correct vacuum of the full theory. Hence the sign fluctuation should be mild around the true vacuum. 
Therefore, if the correct vacuum can be characterized by tuning the value of $\hat{O}$ (e.g. $\hat{O}$ is the pion condensate in QCD$_B$), 
$\langle {\rm e}^{iS_I}\rangle^{(\hat{O})}_i$ and $\langle {\rm e}^{iS_I}\cdot\hat{O}\rangle^{(\hat{O})}_i$ 
can have a sharp peak at a particular value of $i$. 
On the other hand, if the correct vacuum cannot be specified by simply tuning $\hat{O}$, 
their distributions do not show a peak structure and hence the remaining sign problem is not under control. 
 
In summary, the standard density of state method is effective when the quantity of interest characterizes the vacuum of the full theory. 
It is unlikely, however, that a quantity one takes without considering properties of the full theory correctly characterizes its vacuum. 
Therefore there exists a danger that  one has to spend a lot of computational resources to determine a small value of the average for the remaining sign precisely. 
It has been sometimes reported that the remaining sign problem can spoil the density of state method,
due to this inappropriate choice of observables fixed\cite{Anagnostopoulos:2010ux}.

\subsection{Our method}
\hspace{0.51cm}
As we have seen in the previous subsection, 
the traditional density of state method is effective only when the quantity of interest characterizes the vacuum of the full theory, 
since otherwise the overlap problem still exists.   
We solve this problem by slightly changing the viewpoint;  we do not fix the quantity we want to measure. We fix an observable 
which characterizes the correct vacuum, called a good observable
(If more than one quantities  are needed to be specified in order to characterize the vacuum, we  must fix all of them.)   
In the case of the finite-density QCD, pion condensate is such a good observable, 
 since the phase quenched theory becomes exact at large $N_c$ as long as the pion condensate is forbidden by hand\cite{Cherman:2010jj,Hanada:2011ju}. 
Note that we have to understand the physics of the full and phase quenched theories in order to find an appropriate observable 
which characterizes the correct vacuum. 

Let us explain the detail of our idea by using the finite-density QCD. 
We first classify the configurations in the phase-quenched simulation by the values of the pion condensate $\pi^+$. 
(Here $\pi^+$ is defined in terms of QCD$_I$, i.e. the chemical potentials for up and down quarks 
in the operator are $+\mu$ and $-\mu$, respectively, rather than $+\mu$ and $+\mu$.)
Let the height of the histogram at $x_i<\pi^+<x_{i+1}$ be $\rho_i$. 
We also calculate the average sign at $x_i<\pi^+<x_{i+1}$, which we denote $\langle {\rm e}^{iS_I}\rangle_{i}$. 
Then we have a trivial relation, 
\begin{eqnarray}
\langle {\rm e}^{iS_I}\rangle_{\rm P.Q.}
=
\frac{\sum_i \langle {\rm e}^{iS_I}\rangle_i\cdot\rho_i}{\sum_i\rho_i}, 
\end{eqnarray}
where $\rho_i$ is the relative weight factor of $x_i<\pi^+<x_{i+1}$ in the phase quenched simulation\footnote{
Note that we need only {\it relative} weight factor in the region where the phase fluctuation is not very violent. 
Indeed the normalization factor does not play any role in \eqref{eq:reweight}.  
}. 
In the same manner, 
\begin{eqnarray}
\langle {\rm e}^{iS_I}\cdot\hat{O}\rangle_{\rm P.Q.}
=
\frac{\sum_i \langle {\rm e}^{iS_I}\cdot\hat{O}\rangle_i\cdot\rho_i}{\sum_i\rho_i}, 
\end{eqnarray}
where $\hat{O}$ is  an arbitrary operator  we are interested in other than $\pi^+$. 
\begin{figure}[htbp]
   \begin{center}\rotatebox{0}{
   \scalebox{1.5}{
     \includegraphics[height=6.2cm]{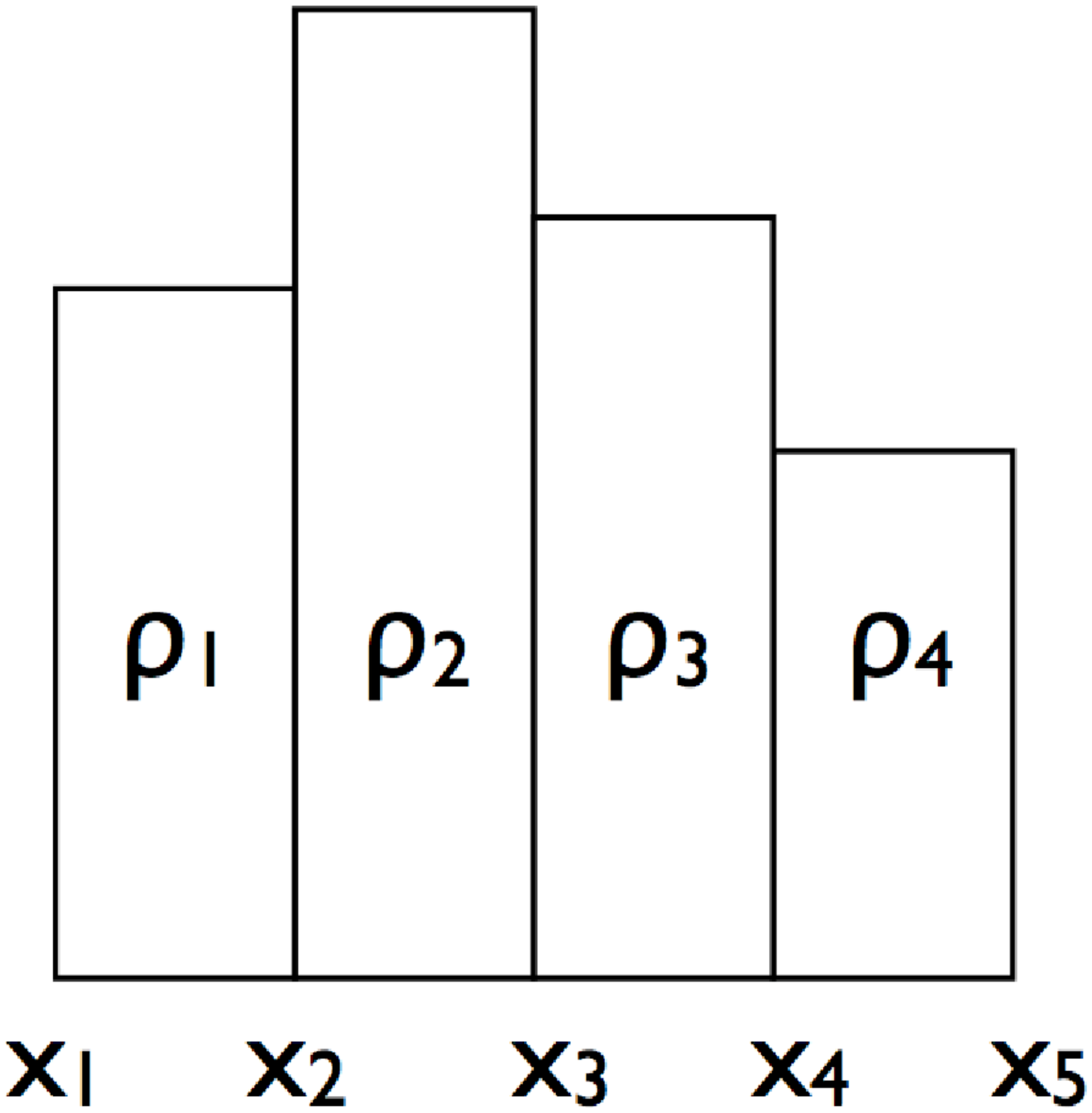}}}
   \end{center}
   \caption{ A histogram of $\pi^+$.}
\label{fig:histogram}
\end{figure} 
Therefore the phase reweighting can be done as 
\begin{eqnarray}
\langle \hat{O}\rangle_{\rm full}
=
\frac{\langle {\rm e}^{iS_I}\cdot\hat{O}\rangle_{\rm P.Q.}}{\langle {\rm e}^{iS_I}\rangle_{\rm P.Q.}}
=
\frac{\sum_i \langle {\rm e}^{iS_I}\cdot\hat{O}\rangle_i\cdot\rho_i}{\sum_i \langle {\rm e}^{iS_I}\rangle_i\cdot\rho_i}.
\label{eq:reweight} 
\end{eqnarray}

We can expect that 
$\sum_i \langle {\rm e}^{iS_I}\rangle_i\cdot\rho_i$ 
and 
$\sum_i \langle {\rm e}^{iS_I}\cdot\hat{O}\rangle_i\cdot\rho_i$ 
takes non-negligible values only around the vacuum of the full theory, $\pi^+=0$. 
Therefore we only have to study there; when 
$\sum_i \langle {\rm e}^{iS_I}\rangle_i\cdot\rho_i$ 
and 
$\sum_i \langle {\rm e}^{iS_I}\cdot\hat{O}\rangle_i\cdot\rho_i$ 
become so small that the precise determination is difficult, 
they do not affect the results and hence we can simply omit them. 
(In fact, unless we have extremely large statistics with which we can determine the small average phase at nonzero $\pi^+$, 
adding such configurations just increases the error. Therefore,  by throwing away un-important configurations 
we can make the result more precise.) 
Note again that this method is not purely numerical;  we know the important samples based on physics. 
The huge sign fluctuation then tells us that we do not have to measure them,  so that it does not increase the simulation cost.  Instead it reduces the cost. 
Therefore the sign problem turns into the sign blessing in this situation.

The actual simulation for RMT goes as follows\footnote{
For QCD, more sophisticated method is needed because the simulation cost is larger. 
See Sec.~\ref{sec:QCD}. 
}. We add a deformation term 
\begin{eqnarray}
\Delta S = \gamma |\pi^+ - x|^2 
\label{deformation_RMT}
\end{eqnarray}
for $ |\pi^+ - x|\ge \epsilon$. The constraint parameter $\gamma$ is taken sufficiently large 
so that all samples lie in $ |\pi^+ - x|< \epsilon$ during the simulation.

Note that there are two options:
\begin{itemize}
\item
Introduce the source both for $S$ and $\Delta S$. 
In this case we have to make the zero source extrapolation in the end. 

\item
Introduce the source only for $\Delta S$. 
In this case we do not need to take the zero source extrapolation.
We take this option in this paper. 

\end{itemize}
It is important to stress that $\pi^+ \ge 0$ as long as $c>0$, so that $\langle \pi^+\rangle \simeq 0$ implies that only $\pi^+\simeq 0$ configurations contribute in the full theory.

\begin{figure}[htbp]
\begin{flushleft}
\hskip 1cm
   \rotatebox{0}{
   \scalebox{0.4}{
     \includegraphics{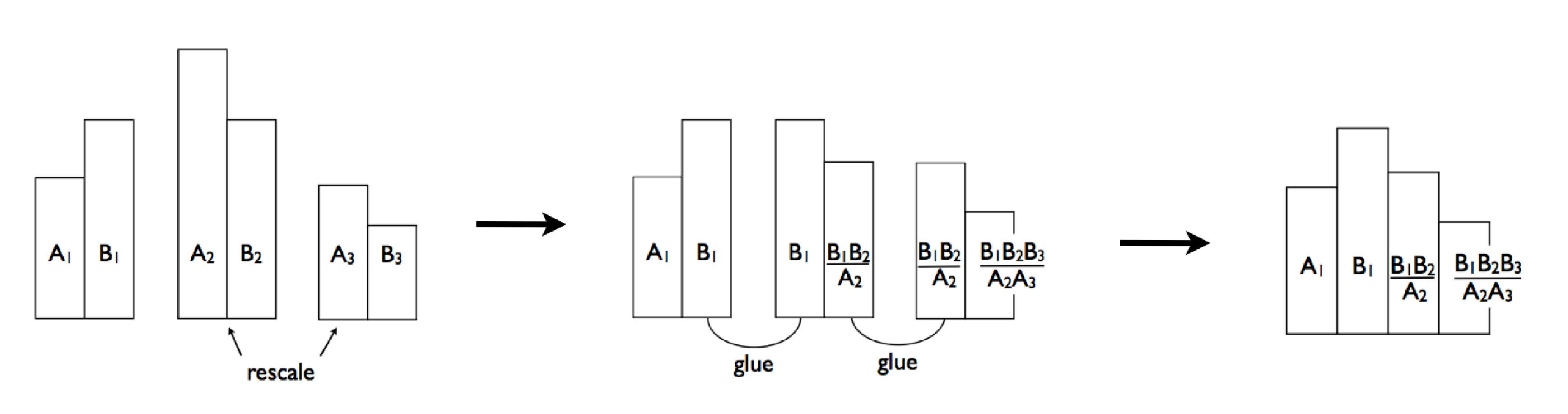}}}
   \end{flushleft}
   \caption{How to obtain the full histogram by gluing partial histograms. }
\label{fig:gluing_histogram}
\end{figure}

Firstly we have to determine the distribution of the histogram of the pion condensate in the phase quenched simulation precisely. 
By introducing the deformation and tuning $x$ and $\epsilon$ we can sample the tail effectively. 
The histograms obtained are `partial' ones restricted at $[x-\epsilon,x+\epsilon]$. 
This situation is like the leftmost panel in Fig.~\ref{fig:gluing_histogram}; 
here the simulation has been done for $x=x_0$ (left), $x=x_0+\epsilon$ (center) and $x=x_0+2\epsilon$, 
with the common value of $\epsilon$, the number of configurations in the partial histograms are 
$A_i$ for $\pi^+<x$ and $B_i$ for $\pi^+>x$. 
In order to obtain the full histogram, we rescale them as in the second panel, 
and then glue them as in the third panel. 

Note that, if the difference of two theories are characterized by many observables, we must fix all of them. 
In the case of finite density QCD, for example, as the isospin chemical potential becomes larger, not just pion but also other fields such as the $\rho$-meson can condense. 
Then we should add deformation terms to fix them.  

Our method could be regarded as an improved version of the multi-parameter factorization method \cite{Anagnostopoulos:2010ux}, 
which has been applied for a supersymmetric matrix model. 
(The `factorization method' is essentially the same as the density of state method.) 
For this improvement, a good understanding about the physics under consideration is crucial. 
In the multiple-parameter factorization method, one labels the configurations by values of a set of multiple observables, 
$\hat{O}_1, \hat{O}_2, \cdots, \hat{O}_n$, and 
\begin{eqnarray}
\langle {\rm e}^{iS_I}\rangle_{P.Q.}
=
\frac{
\sum_{i_1,i_2, \cdots, i_n} \langle {\rm e}^{iS_I}\rangle^{(\hat{O}_1,\cdots,\hat{O}_n)}_{i_1,i_2, \cdots, i_n}\cdot\rho^{(\hat{O}_{1},\cdots,\hat{O}_n)}_{i_1,i_2, \cdots, i_n}
}{
\sum_{i_1,i_2, \cdots, i_n}\rho^{(\hat{O}_{1},\cdots,\hat{O}_n)}_{i_1,i_2, \cdots, i_n}
},   
\end{eqnarray}
and similarly for $\langle \hat{O}\rangle_{\rm full}$. 
One can expect that $\langle {\rm e}^{iS_I}\rangle^{(\hat{O}_1,\cdots,\hat{O}_n)}_{i_1,i_2, \cdots, i_n}$ has a single peak 
by introducing sufficiently many observables. 
In other words, the overlap problem can be solved by fixing sufficiently many observables. 
Suppose the quantity in consideration, say $\hat{O}_1$, does not characterize the vacuum. Then the overlap problem is not solved. 
In the case of QCD, the overlap problem can be solved by taking $\hat{O}_2$ to be the pion condensate. 
(When necessary one should also add $\rho$-condensate as $\hat{O}_3$ etc.) 
But then we do not even have to fix $\hat{O}_1$, because it is not the source of the overlap problem anyways.  
Then, by letting $\hat{O}_1$ take any value, we arrive at our method. 
The point is that we only have to fix the quantities characterizing the correct vacuum, and for that purpose we have to understand 
the difference between physics of full and phase quenched theories. 
For that, nonperturbative arguments like the large $N_c$ equivalence \cite{Cherman:2010jj,Hanada:2011ju} play important roles. 
(In the supersymmetric matrix model studied in \cite{Anagnostopoulos:2010ux}, 
they identified the observables which characterize the vacuum by using another numerical method, 
and then applied the multi-parameter factorization method.) 
Whether one has to fix multiple observables or not is a problem-specific issue, which depends on theories and parameter regions.

A good understanding about the vacuum structure of the full and phase-quenched theories is very important 
for this method to work. In the case of QCD, we already know the right quantity to fix. 
(We could also choose other quantities, but then we would have to fix multiple quantities, 
which makes actual calculation more difficult.) 
We know that only $\pi^+\sim 0$ is important, and can safely neglect the parameter region with small average sign. 
We do not even have to study large $\pi^+$ region.  
If we didn't know the right quantity to fix, we would have to measure the small sign rather precisely, 
in order to make sure that such parameter region is not important. 

If we consider other theories for which the physical interpretation of the phase quenched theory is not clear, 
the simplest way to find `good' observables would be to calculate various observables in the phase-quenched theory 
whose counterparts in the full theory trivially vanish due to symmetries. 
In case  that the full theory is not understood well, one has to try purely numerical method: 
fix various quantities, scan the parameter space and find a nice peak structure of the average phase, 
as suggested in \cite{Anagnostopoulos:2010ux}. 

\subsubsection{A comment on `silver-blaze' region $\mu<\mu_c$}\label{sec:comment_baryon_pin_dome}
\hspace{0.51cm}
In QCD$_B$, at zero temperature and at the `silver blaze' region $\mu<\mu_c$, $\nu_B$ must be zero.
Therefore, at $\mu<\mu_c$, it is possible to reduce the overlapping problem by setting $\nu_B$, rather than the pion condensate, to zero. 
To demonstrate this in the RMT, however, we should take $\nu_B$ to be some negative value, since
$\nu_B$ becomes negative in this region due to an RMT-artifact. 
In the RMT$_I$, the observable $\nu_B$ corresponds to the isospin density $\nu_I$, which is positive. 
In Sec.~\ref{sec:fix_nu}, we also consider the behavior of the average phase as a function of $\nu_B$. 

\section{Simulation results}\label{sec:simulation}
\hspace{0.51cm}

\begin{figure}[htbp]
   \begin{center}\rotatebox{0}{
   \scalebox{0.7}{
     \includegraphics[height=6cm]{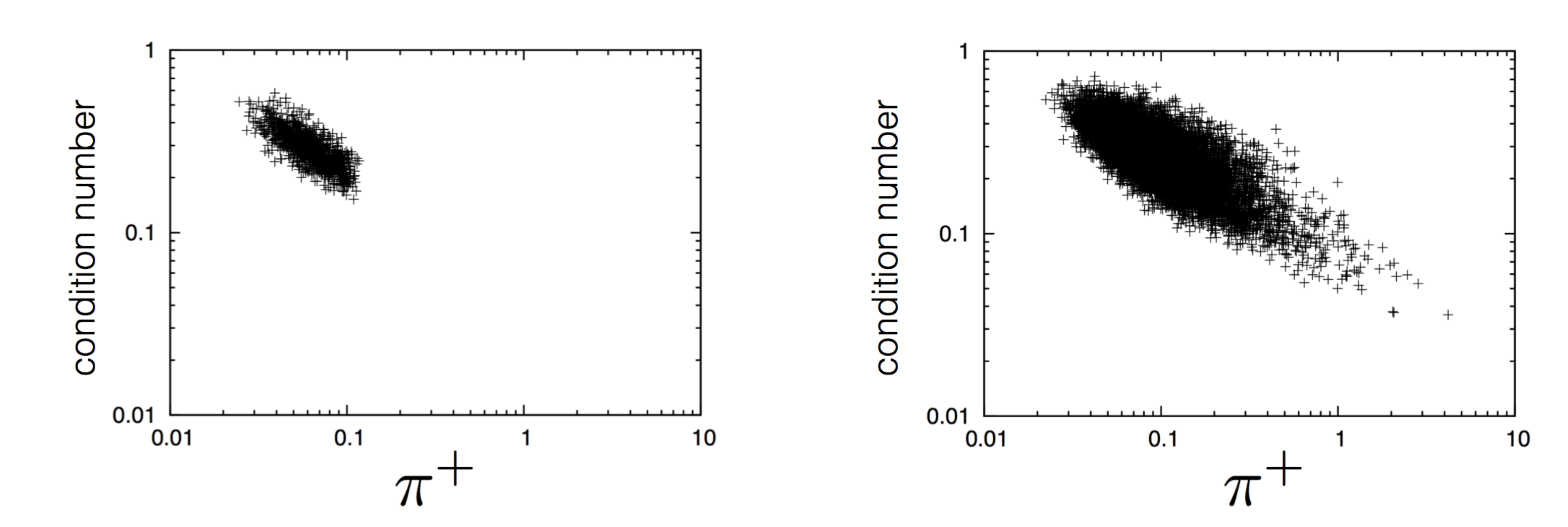}}}
   \end{center}
  \caption{Distribution of the condition number ($|{\rm (maximum\ eigenvalue)}/{\rm (minimum\ eigenvalue)}|$) and $\pi^+$.
  $N=4$, $m=0.35$, $\mu=0$ (left) and $\mu=0.7$ (right).}\label{fig:N4M035_condition_number}
\end{figure}

In this section we show the simulation results of 
$N_f=2$ RMT.
In order to see the effect of the pion condensate, let us start with a nonzero mass. 
(When mass is zero, $\mu=0$ is already at the border of the pion condensation.)
In Fig.~\ref{fig:N4M035_condition_number} we show the distribution of the condition number, 
$|{\rm (minimum\ eigenvalue\ of}\ {\cal D}_f)/{\rm (maximum\ eigenvalue\ of}\ {\cal D}_f)|$, and the pion condensate $\pi^+$, 
for $N=4$, $m=0.35$, and $c=0.02$ , $\mu=0$ and $\mu=0.7$. (Note that the source $c$ is introduced only for the constraint term $\Delta S$.) 
At $\mu=0$, pion does not condense, and hence $\pi^+$ takes small values. 
At $\mu=0.7$, the distribution of $\pi^+$ has a long tail, which is the signature of the pion condensation. 
We can see that the condition number becomes smaller as $\pi^+$ increases, 
as expected from the fact that the pion condensate is caused by near zero-mode 
in the Dirac spectrum. 
 
\begin{figure}[htbp]
   \begin{center}\rotatebox{0}{
   \scalebox{1.2}{
     \includegraphics[height=6.5cm]{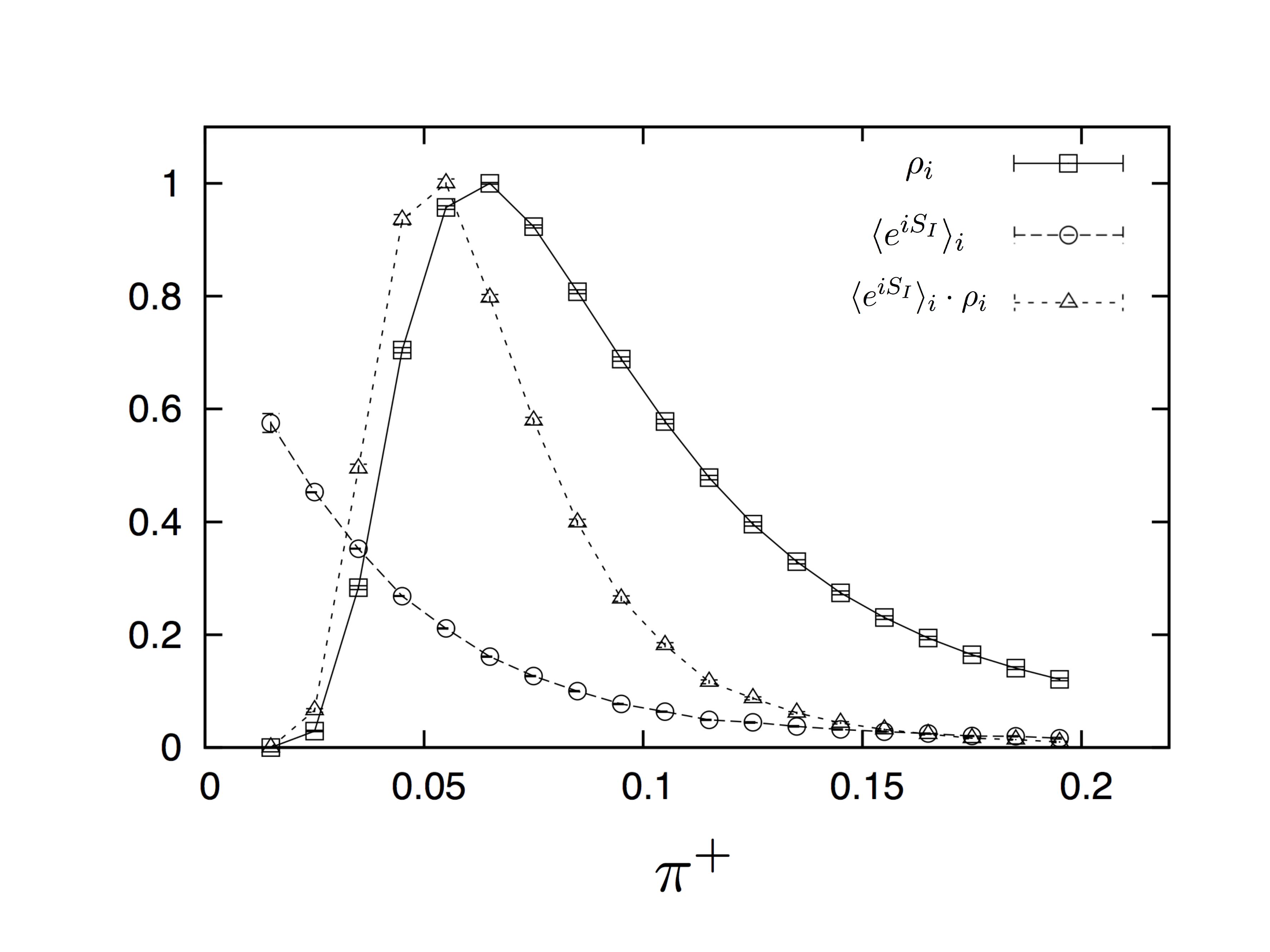}}}
   \end{center}
   \caption{The average phase $\langle {\rm e}^{iS_I}\rangle_i$ and relative wight with and without phase, $\rho_i$ and $\langle {\rm e}^{iS_I}\rangle_i\cdot \rho_i$. 
   $N=4$, $m=0.35$, $\mu=0.7$ and $c=0.02$. 
   The peaks of $\rho_i $ and $\rho_i \cdot \langle {\rm e}^{iS_I}\rangle_i$ are normalized to be 1. 
   }
\label{fig:relative_weight_N4M035C070}
\end{figure}

\begin{figure}[htbp]
   \begin{center}\rotatebox{0}{
   \scalebox{1.2}{
     \includegraphics[height=6.5cm]{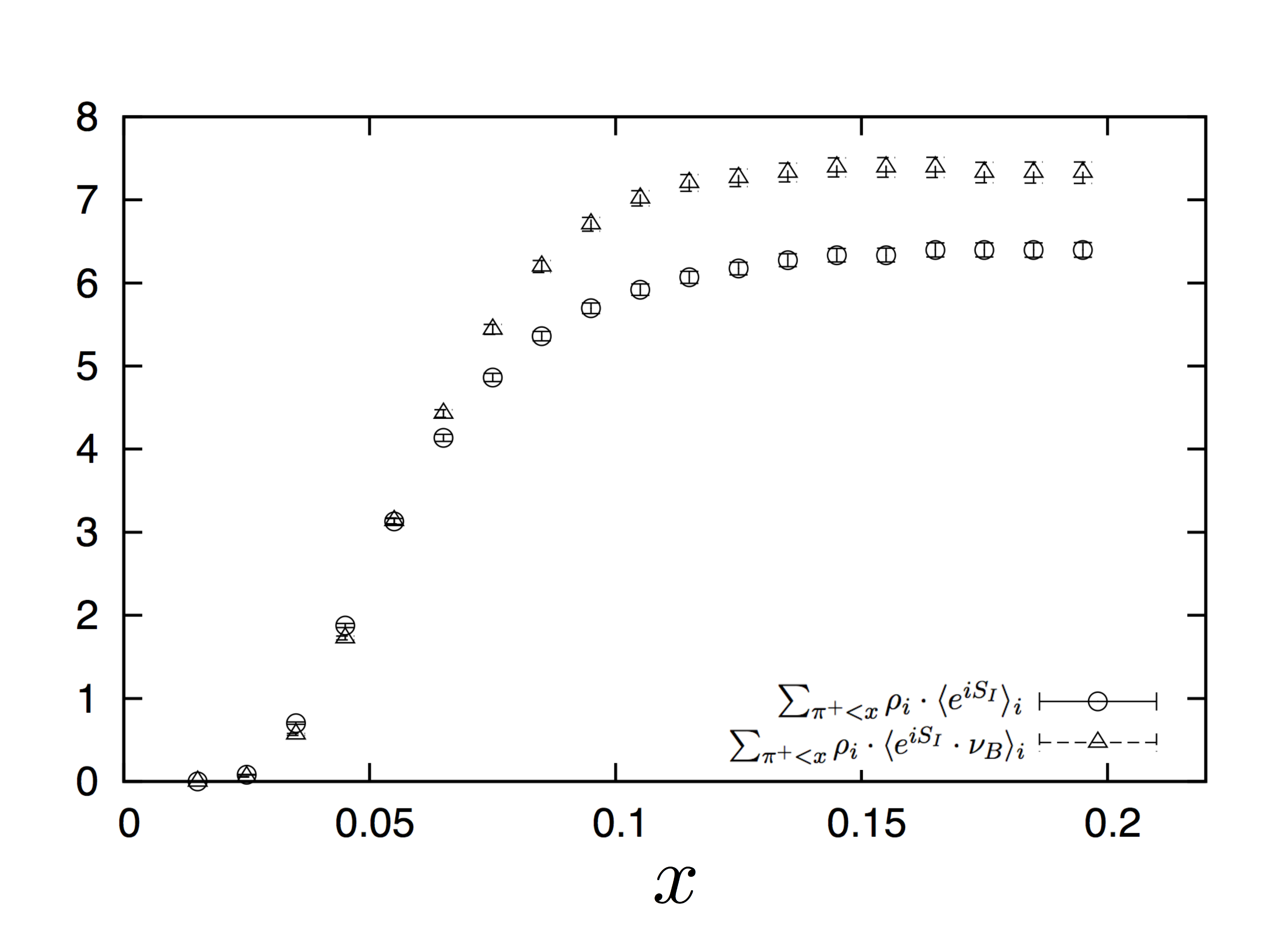}}}
   \end{center}
   \caption{ $\sum_{\pi^+<x} \rho_i\cdot \langle {\rm e}^{iS_I}\rangle_i$ and $\sum_{\pi^+<x} \rho_i \cdot \langle {\rm e}^{iS_I}\cdot \nu_B\rangle_i$ calculated at limited range of $\pi^+$.
   $N=4$, $m=0.35$, $\mu=0.7$ and $c=0.02$.  
   The normalization is the same as in Fig.~\ref{fig:relative_weight_N4M035C070}, 
    i.e. the peak of $\rho_i \cdot\langle {\rm e}^{iS_I}\rangle_i$ is normalized to be 1. 
   }
\label{fig:N4M035C070_phase_sum_convergence}
\end{figure}

In Fig.~\ref{fig:relative_weight_N4M035C070} we show the average phase $\langle {\rm e}^{iS_I}\rangle_i$, 
relative weight $\rho_i$, and the reweighted relative weight $\rho_i\cdot\langle {\rm e}^{iS_I}\rangle_i$ 
for $\mu=0.7$. 
The weight in the phase-quenched simulation has a long tail reflecting the pion condensation.  
The average phase becomes small as $\pi^+$ becomes large, so that this fat tail is removed in the reweighted relative weight $\rho_i\cdot\langle {\rm e}^{iS_I}\rangle_i$.
Therefore the large-$\pi^+$ region gives only a negligible contribution;  
in fact, as shown in Fig.~\ref{fig:N4M035C070_phase_sum_convergence}, 
$\sum_{\pi^+<x} \rho_i\cdot \langle {\rm e}^{iS_I}\rangle_i$ and $\sum_{\pi^+<x} \rho_i \cdot \langle {\rm e}^{iS_I}\cdot \nu_B\rangle_i$ 
calculated at limited range of $\pi^+$ quickly converges. We can terminate the sum at around $\pi^+\sim 0.15$, so that 
we will not see small condition numbers. 


\vskip 0.5 cm

Next let us consider the massless limit, where the sign problem is severe.  
At each $N$, the value of the baryon density $\nu_B$ can be calculated analytically\cite{Halasz:1997he}. 
For example, 
\begin{eqnarray}
\nu_B
=
\frac{
-180 \mu + 1440 \mu^3 - 5760 \mu^5 + 15360 \mu^7 -24960 \mu^9 +
    24576 \mu^{11} - 14336 \mu^{13} +
    4096 \mu^{15}}{45 - 360 \mu^2 + 1440 \mu^4 - 3840 \mu^6 + 7680 \mu^8 -
      9984 \mu^{10} + 8192 \mu^{12} - 4096 \mu^{14} + 1024 \mu^{16}}
\end{eqnarray}
for $N=4$. 

\begin{figure}[htbp]
   \begin{center}\rotatebox{0}{
   \scalebox{1.2}{
     \includegraphics[height=6.5cm]{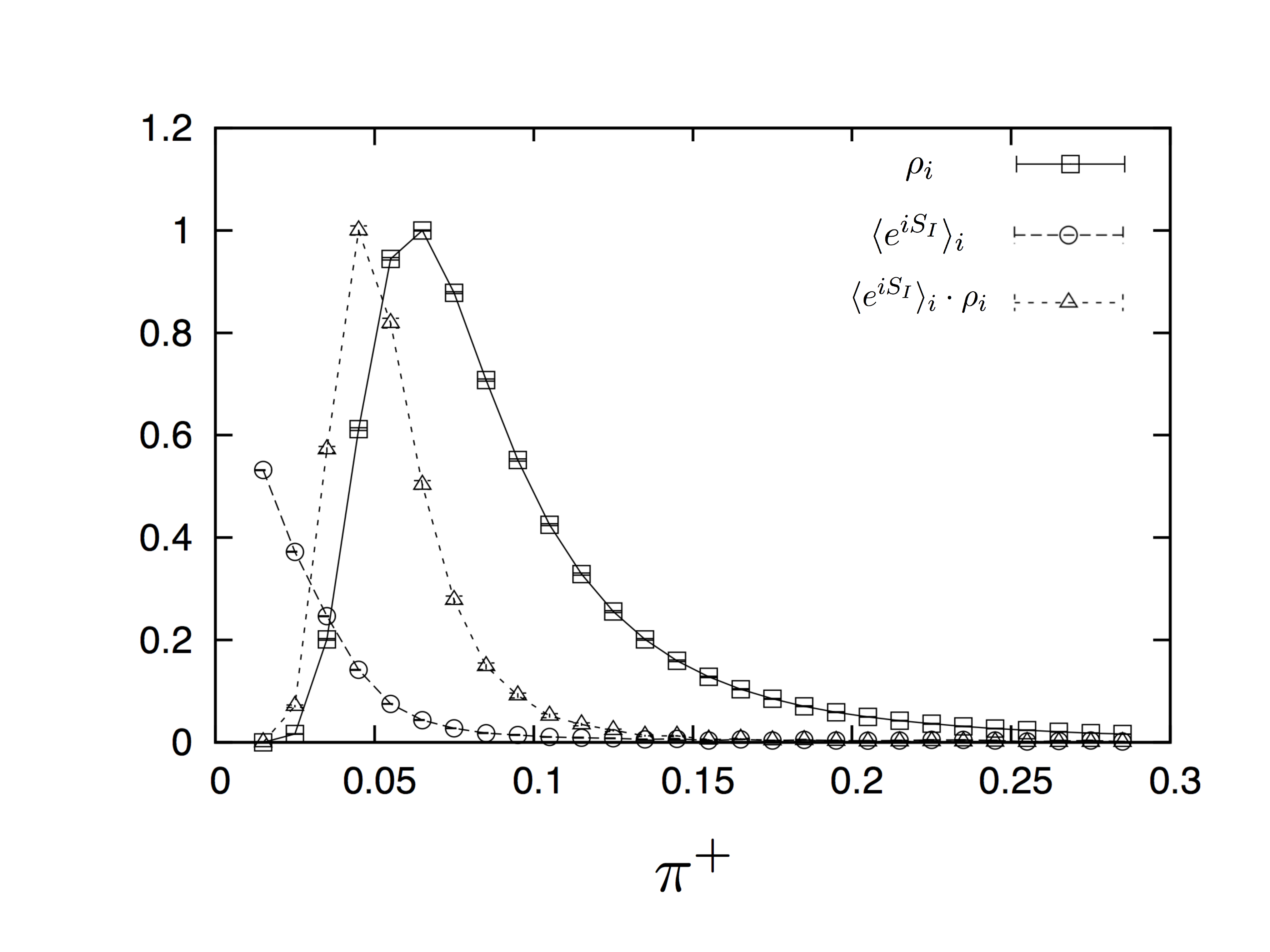}}}
   \end{center}
   \caption{The average phase $\langle {\rm e}^{iS_I}\rangle_i$ and relative wight with and without phase, $\rho_i$ and $\langle {\rm e}^{iS_I}\rangle_i\cdot \rho_i$. 
   $N=4$, $m=0$, $\mu=0.7$ and $c=0.02$. 
   The peaks of $\rho_i $ and $\rho_i \cdot \langle {\rm e}^{iS_I}\rangle_i$ are normalized to be 1. 
   }
\label{fig:relative_weight_N4M000C070}
\end{figure}

\begin{figure}[htbp]
   \begin{center}\rotatebox{0}{
   \scalebox{1.2}{
     \includegraphics[height=6.5cm]{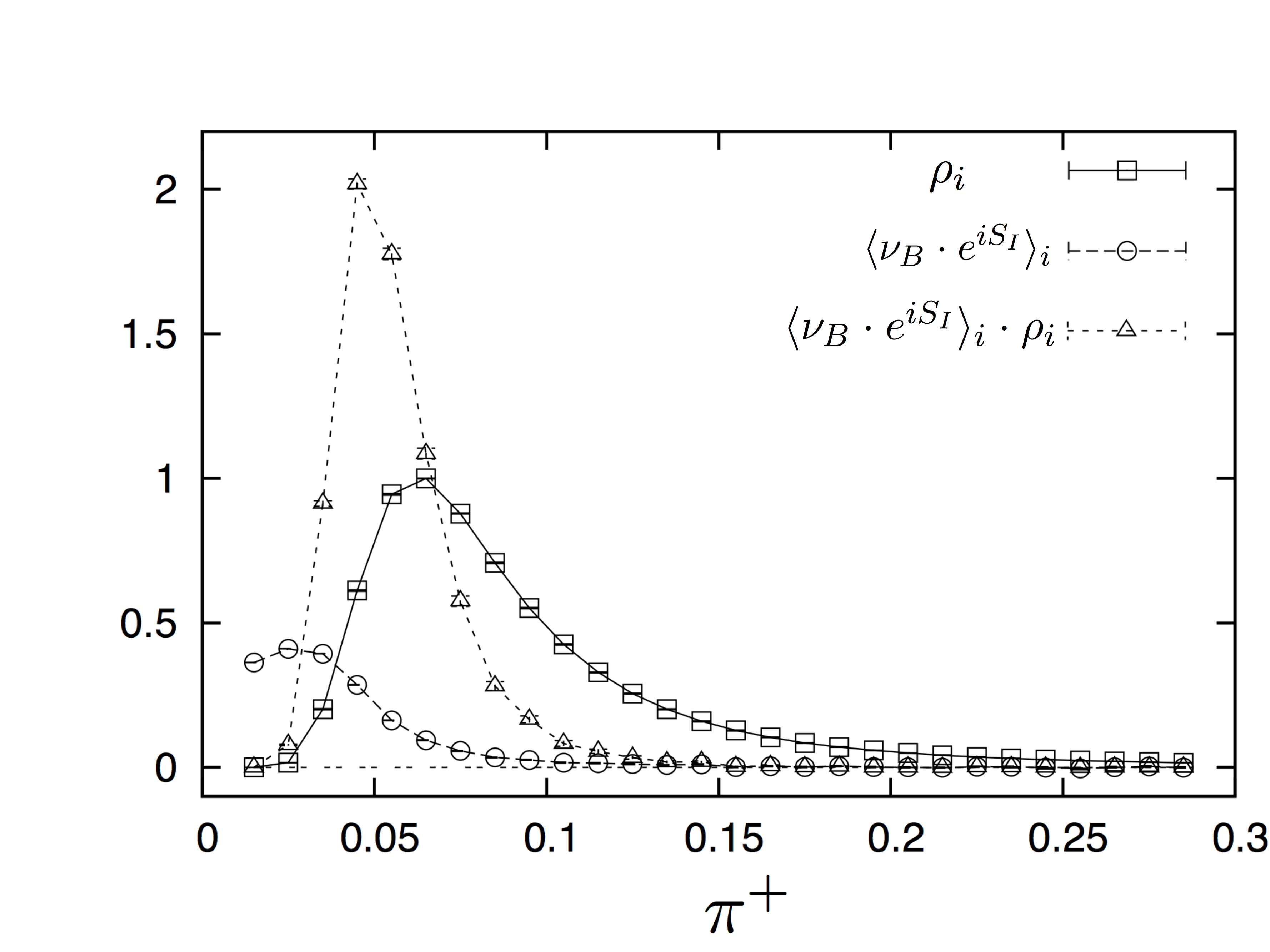}}}
   \end{center}
   \caption{$\langle \nu_B\cdot {\rm e}^{iS_I}\rangle_i$ and relative wight with and without phase, $\rho_i$ and $\langle \nu_B\cdot  {\rm e}^{iS_I}\rangle_i\cdot \rho_i$. 
   $N=4$, $m=0$, $\mu=0.7$ and $c=0.02$.    The normalization is the same as in Fig.~\ref{fig:relative_weight_N4M000C070}. }
\label{fig:BaryonDensity_N4M000C070}
\end{figure}

Let us consider $N=4$, $\mu=0.7$ and $c=0.02$ as an example. 
We take $\epsilon=0.01$, $x=0.01, 0.02, 0.03,\cdots$. 
For each bin, we collected $1,000,000$ configurations. 
In Fig.~\ref{fig:relative_weight_N4M000C070} we show the average phase $\langle {\rm e}^{iS_I}\rangle_i$, 
relative weight $\rho_i$, and the reweighted relative weight $\rho_i\cdot\langle {\rm e}^{iS_I}\rangle_i$. 
The weight in the phase-quenched simulation has a long tail reflecting the pion condensation. 
However the average phase is extremely small at this tail and the reweighted relative weight does not have a fat tail. 
Also the peak is shifted to a small-$\pi^+$ region. 
%
In Fig.~\ref{fig:BaryonDensity_N4M000C070}, $\rho_i\cdot\langle {\rm e}^{iS_I}\cdot\nu_B\rangle_i$ and 
$\langle {\rm e}^{iS_I}\cdot\nu_B\rangle_i$ are plotted. 
$\langle {\rm e}^{iS_I}\cdot\nu_B\rangle_i$ behaves similarly to $\langle {\rm e}^{iS_I}\rangle_i$: it approaches zero very quickly. 
As a result, $\rho_i\cdot\langle {\rm e}^{iS_I}\cdot\nu_B\rangle_i$ does not have a fat tail either. 
%

\begin{figure}[htbp]
   \begin{center}\rotatebox{0}{
   \scalebox{1.2}{
     \includegraphics[height=6.5cm]{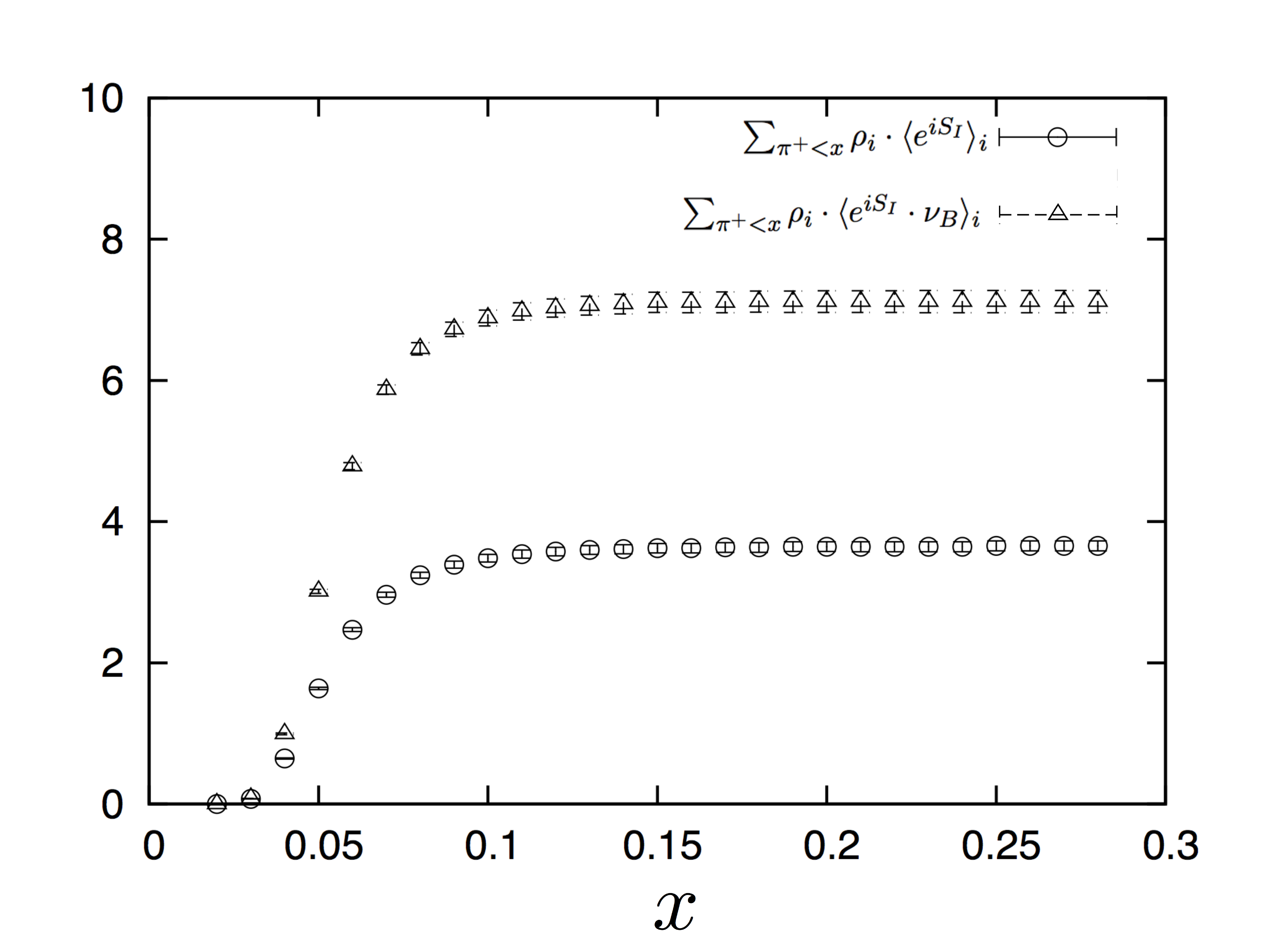}}}
   \end{center}
   \caption{ $\sum_{\pi^+<x} \rho_i\cdot \langle {\rm e}^{iS_I}\rangle_i$ and $\sum_{\pi^+<x} \rho_i \cdot \langle {\rm e}^{iS_I}\cdot \nu_B\rangle_i$ calculated at limited range of $\pi^+$. 
   $N=4$, $m=0$, $\mu=0.7$ and $c=0.02$.  
   The normalization is the same as in Fig.~\ref{fig:relative_weight_N4M000C070} 
   and Fig.~\ref{fig:BaryonDensity_N4M000C070}, i.e. the peak of $\rho_i \cdot\langle {\rm e}^{iS_I}\rangle_i$ is normalized to be 1. 
   }
\label{fig:N4M000C070_phase_sum_convergence}
\end{figure}

\begin{figure}[htbp]
   \begin{center}\rotatebox{0}{
   \scalebox{1.2}{
     \includegraphics[height=6.5cm]{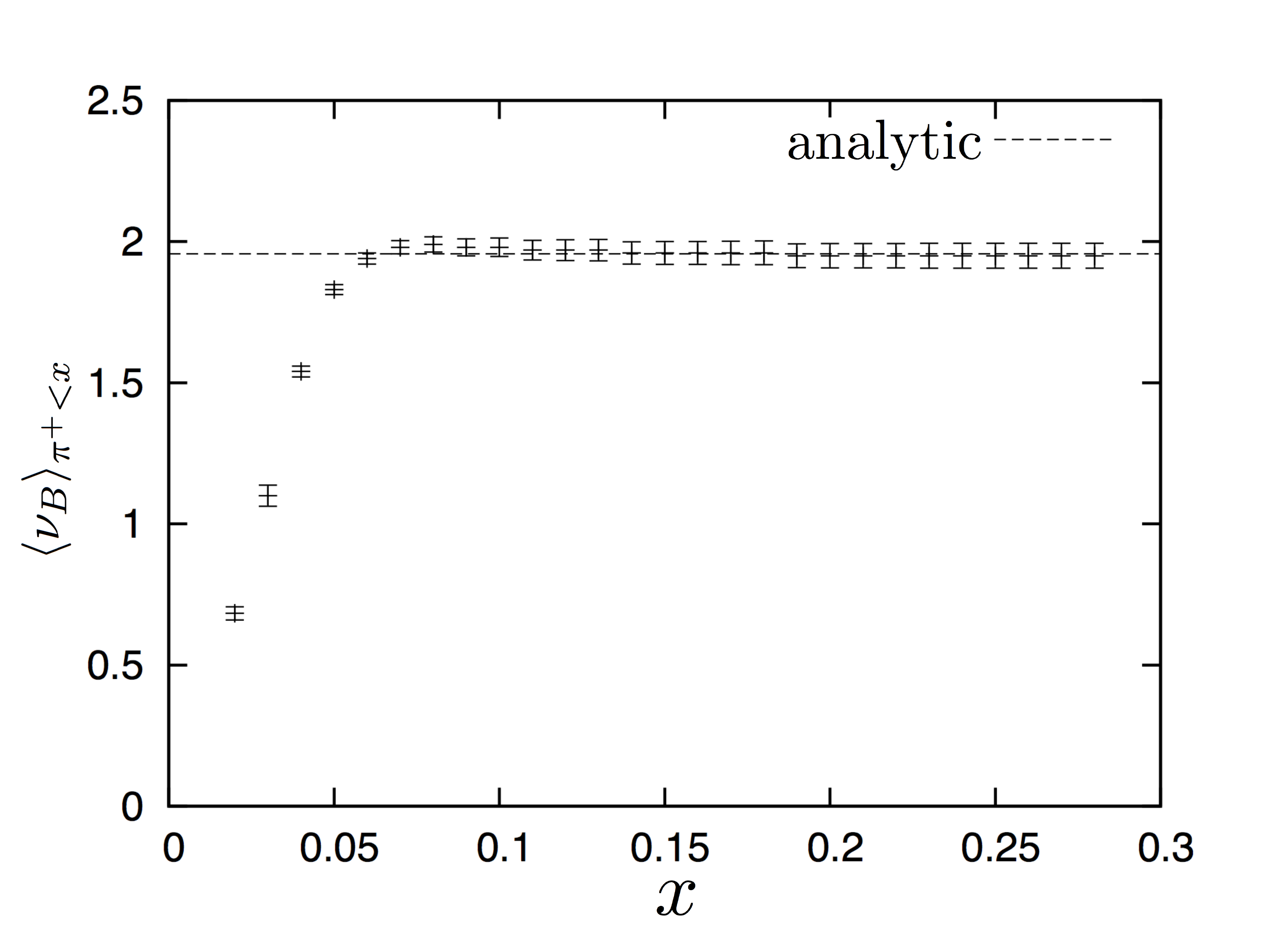}}}
   \end{center}
   \caption{$\langle \nu_B\rangle_{\pi^+<x}$ calculated at limited range of $\pi^+$. 
   $N=4$, $m=0$, $\mu=0.7$ and $c=0.02$. We can see a nice convergence to the exact analytic value. }
\label{fig:BaryonDensity_reweighted_0.7}
\end{figure}

From Fig.~\ref{fig:relative_weight_N4M000C070} and Fig.~\ref{fig:BaryonDensity_N4M000C070}, we can see that $\pi^+\gtrsim 0.08$ is negligible.  
It can be explicitly seen from $\sum_{\pi^+<x} \rho_i \langle {\rm e}^{iS_I}\rangle_i$ and $\sum_{\pi^+<x} \rho_i \cdot \langle {\rm e}^{iS_I}\cdot \nu_B\rangle_i$ 
shown in Fig.~\ref{fig:N4M000C070_phase_sum_convergence}.  
In Fig.~\ref{fig:BaryonDensity_reweighted_0.7}, we plot $\nu_B|_{\pi^+<x}$ as a function of $x$. We can see a good convergence to the analytic value. 
In a usual phase-reweighting method, most computational resources are wasted to evaluate a very small average sign at $\pi^+\gtrsim 0.08$, 
in order to prove that this region is not important. But from the beginning, we knew it is irrelevant. Then why do we have to waste resources there ?

\begin{figure}[htbp]
   \begin{center}\rotatebox{0}{
   \scalebox{1.2}{
     \includegraphics[height=6.5cm]{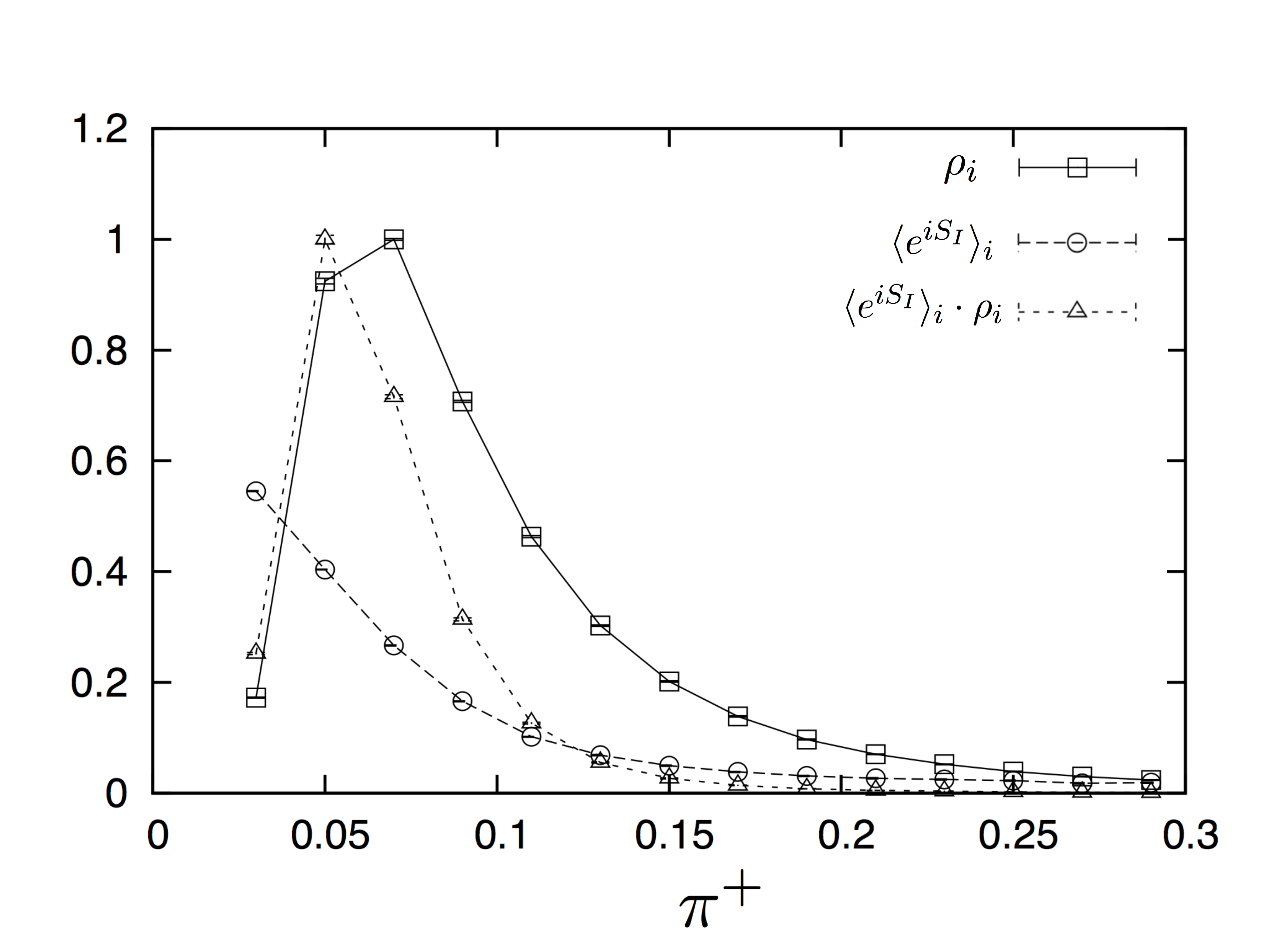}}}
   \end{center}
   \caption{The average phase $\langle {\rm e}^{iS_I}\rangle_i$ and relative wight with and without phase, $\rho_i$ and $\rho_i\cdot\langle {\rm e}^{iS_I}\rangle_i$. 
   $N=4$, $m=0$, $\mu=0.4$ and $c=0.02$. 
   $\langle {\rm e}^{iS_I}\rangle_i$ seems to have a long tail (though the value is not very large). 
   }
\label{fig:relative_weight_N4M000C040}
\end{figure}

\begin{figure}[htbp]
   \begin{center}\rotatebox{0}{
   \scalebox{1.2}{
     \includegraphics[height=6.5cm]{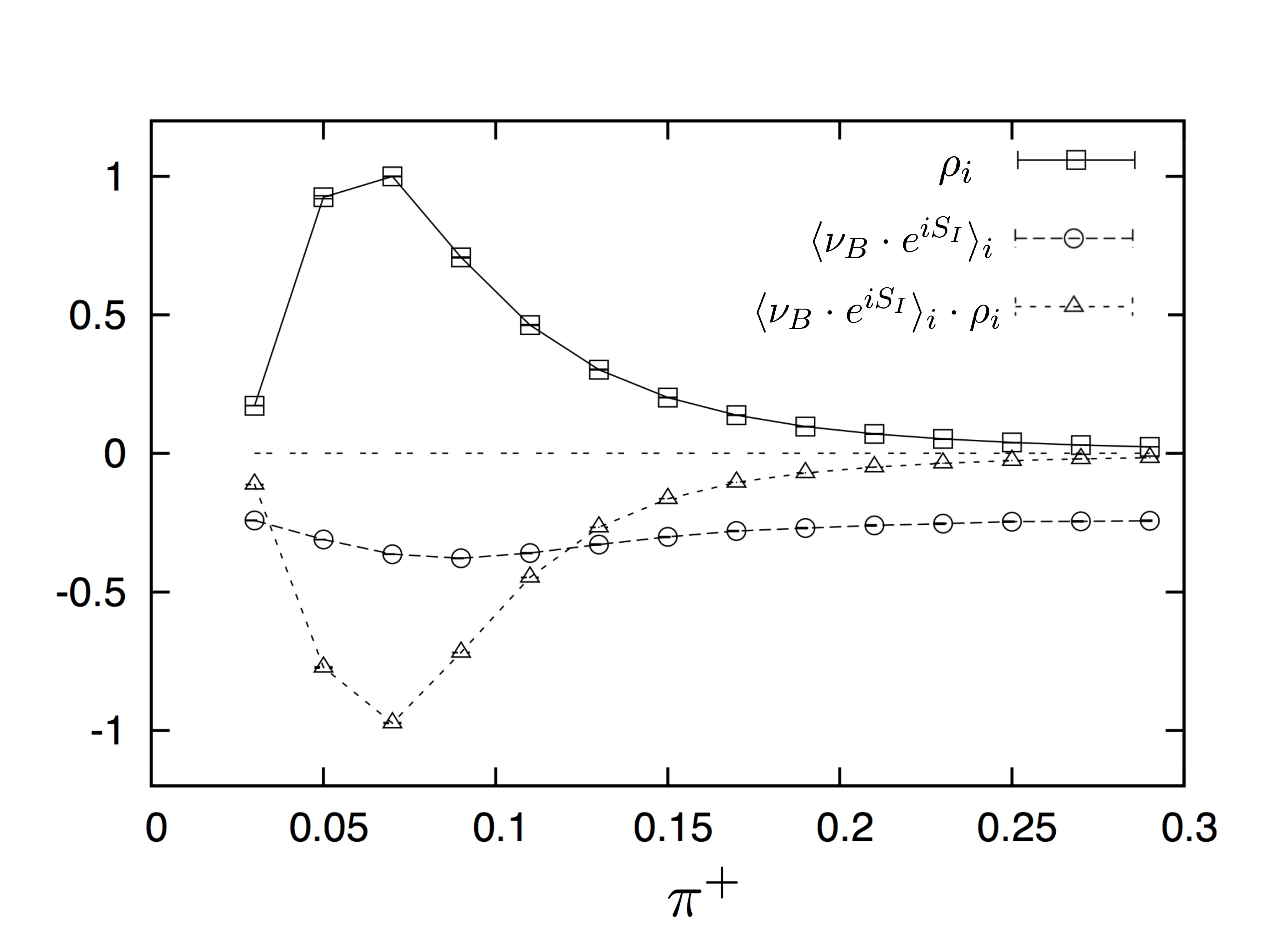}}}
   \end{center}
   \caption{ $\langle \nu_B\cdot {\rm e}^{iS_I}\rangle_i$ and relative wight with and without phase, $\rho_i$ and $\rho_i\cdot\langle  \nu_B\cdot {\rm e}^{iS_I}\rangle_i$. 
   $N=4$, $m=0$, $\mu=0.4$ and $c=0.02$. Although $\langle \nu_B\cdot {\rm e}^{iS_I}\rangle_i$ seems to take a nonzero value at large $\pi^+$, $\rho_i\cdot\langle \nu_B\cdot {\rm e}^{iS_I}\rangle_i$ 
   goes to zero rather quickly because $\rho_i$ becomes zero. 
   Note that the baryon density takes a negative value here because of an artifact of RMT.}
\label{fig:BaryonDensity_N4M000C040}
\end{figure}

Let us also see the plots at $\mu=0.4$, which is below $\mu_c$. 
As shown in Fig.\ref{fig:relative_weight_N4M000C040}, the average phase, though small, seems to remain finite. 
However  $\rho_i$ in the phase-quenched ensemble approaches zero faster than at $\mu=0.7$  at large $\pi^+$, and the distribution after the reweighting is 
similar to that at $\mu=0.7$. 
As for the baryon density, the behavior is very different from the counterpart at $\mu=0.7$. 
As shown in Fig.~\ref{fig:BaryonDensity_N4M000C040}, $\langle \nu_B\,  {\rm e}^{iS_I}\rangle_i$ seems to take a nonzero value at large $\pi^+$. 
However $\rho_i\cdot\langle \nu_B\, {\rm e}^{iS_I}\rangle_i$ goes to zero rather quickly because $\rho_i$ becomes zero. 
(Note that the baryon density takes a negative value here because of an artifact of RMT.)

\begin{figure}[htbp]
   \begin{center}\rotatebox{0}{
   \scalebox{1.2}{
     \includegraphics[height=6.5cm]{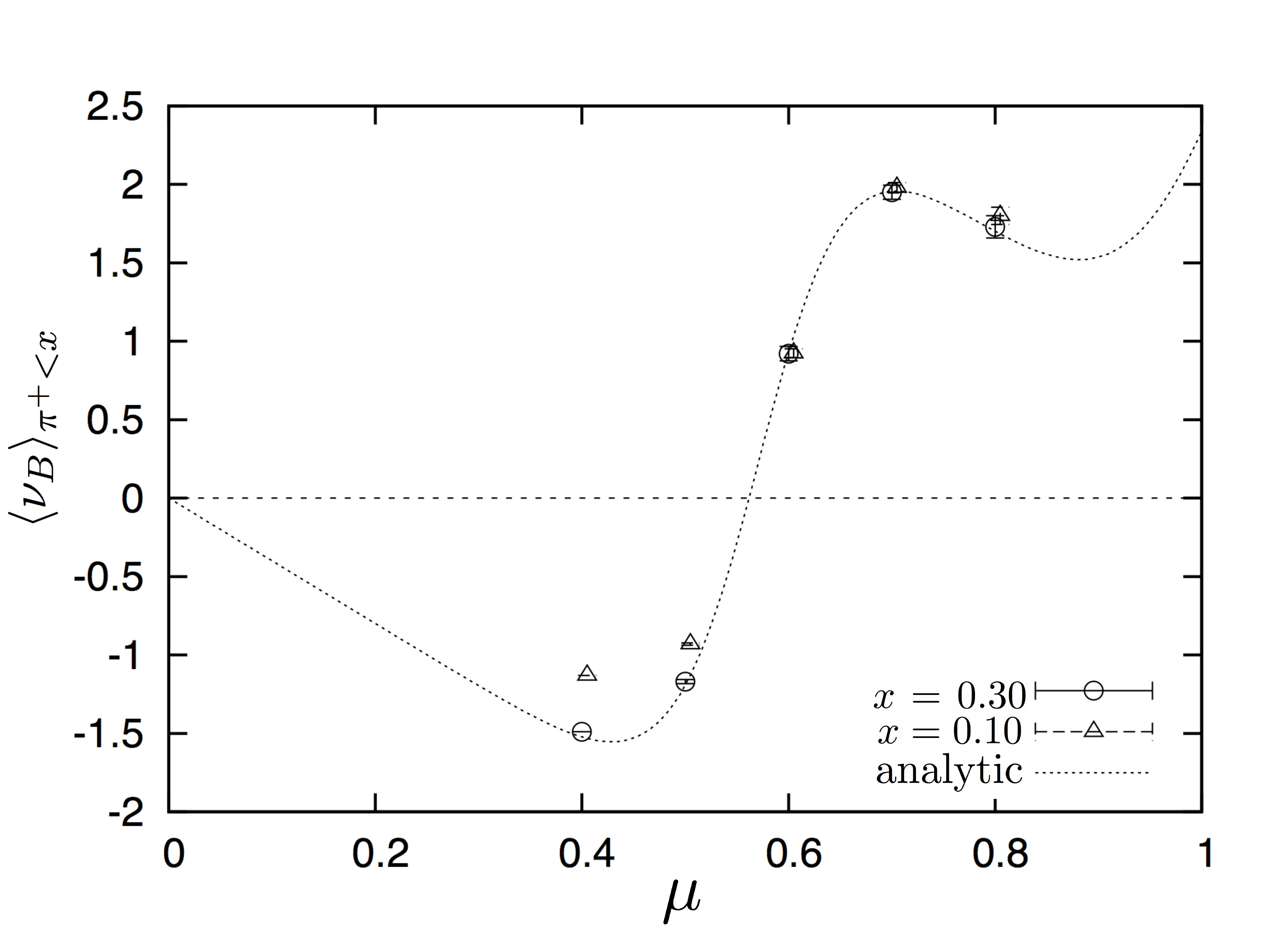}}}
   \end{center}
   \caption{$N=4$, $m=0$ and $c=0.02$, exact value vs. $\langle \nu_B\rangle_{\pi^+<x}$, $x=0.10$ and $x=0.30$ for several values of $\mu$. 
   Data points for $x=0.10$ are shifted to $x$-direction slightly so that they do not overlap with those for $x=0.30$. 
   The convergence to the exact value is slower at $\mu<\mu_c$, because of a fatter tail. 
   }
\label{fig:N4M000_pin_dome_convergence_various_mu}
\end{figure}

In Fig.~\ref{fig:N4M000_pin_dome_convergence_various_mu},
we compare our results of $\langle \nu_B\rangle_{\pi^+<x}$ at $x=0.10$ and $0.30$ with exact results at several values of $\mu$. From this figure we conclude that our method reproduces exact results quite well, though convergences are slower at $\mu < \mu_c$.

\begin{figure}[htbp]
   \begin{center}\rotatebox{0}{
   \scalebox{1.5}{
     \includegraphics[height=6.2cm]{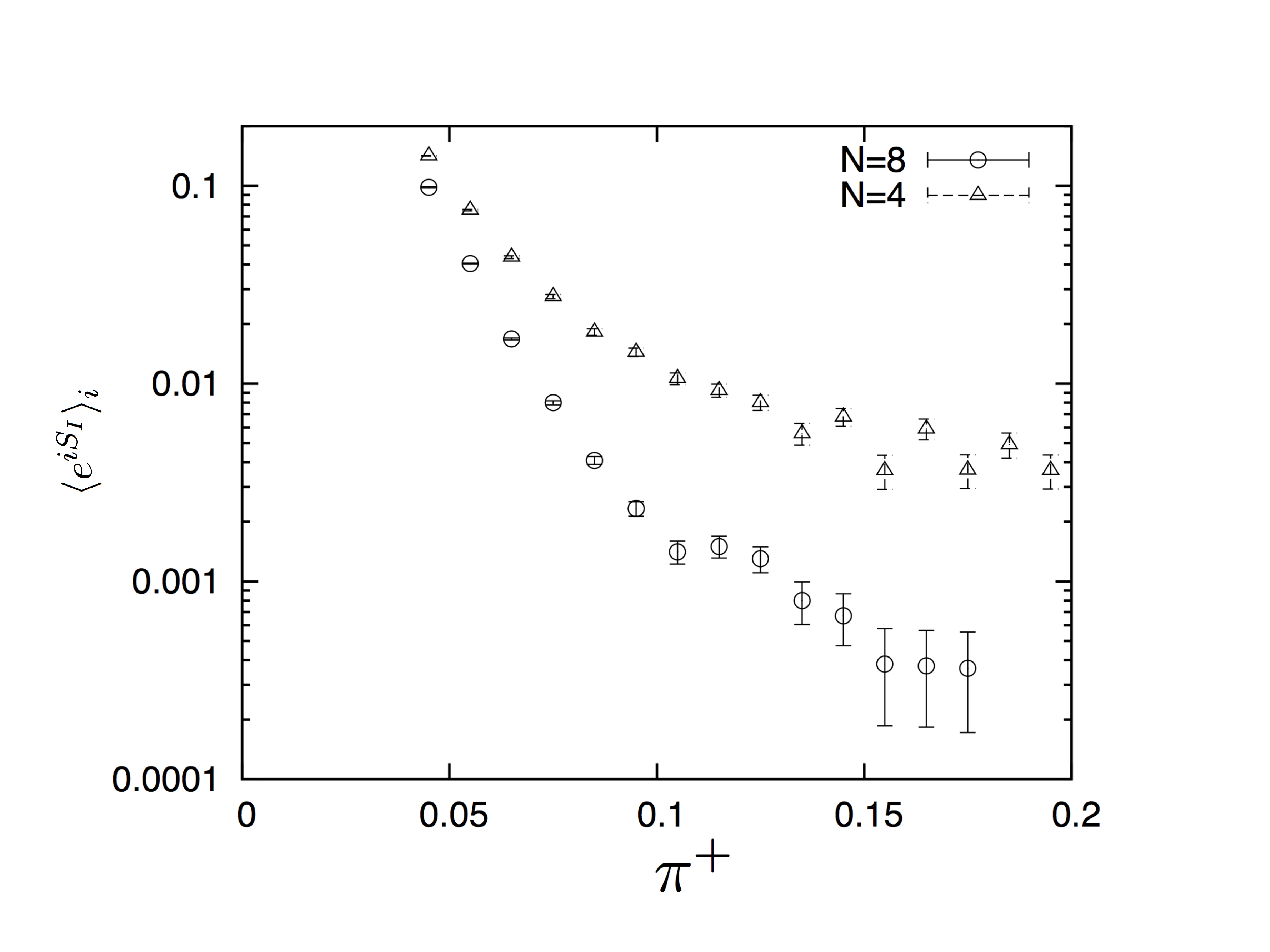}}}
   \end{center}
   \caption{ $\langle {\rm e}^{iS_I}\rangle_i$. 
   $N=4$ and $N=8$, $m=0$, $\mu=0.7$ and $c=0.02$. 
   }
\label{fig:N8M000C070_phase}
\end{figure}

\begin{figure}[htbp]
   \begin{center}\rotatebox{0}{
   \scalebox{1.5}{
     \includegraphics[height=6.2cm]{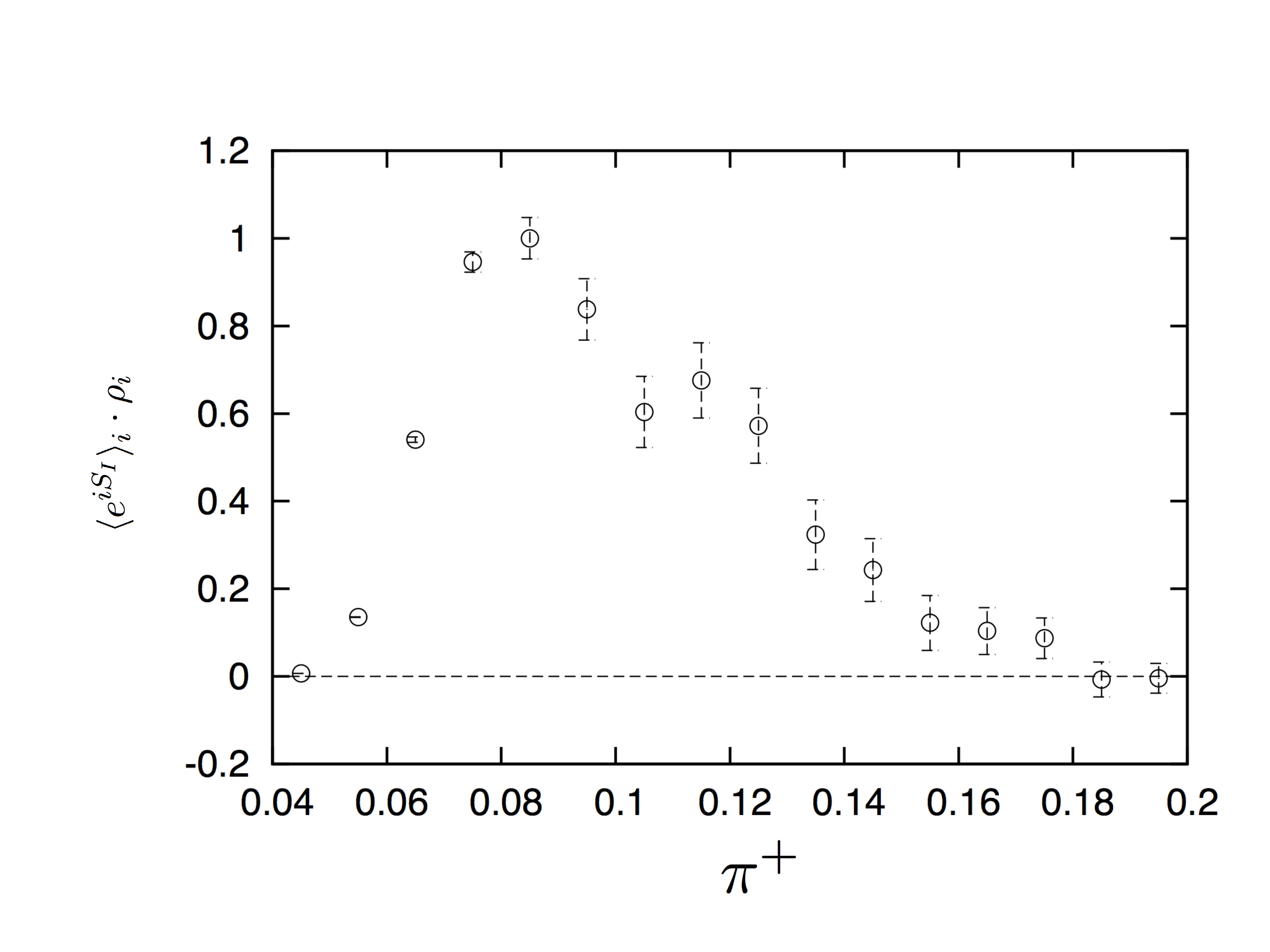}}}
   \end{center}
   \caption{ $\rho_i\cdot\langle {\rm e}^{iS_I}\rangle_i$. 
   $N=8$, $m=0$, $\mu=0.7$ and $c=0.02$. 
   }
\label{fig:N8M000C070S002RelativeWeight}
\end{figure}

\begin{figure}[htbp]
   \begin{center}\rotatebox{0}{
   \scalebox{1.5}{
     \includegraphics[height=6cm]{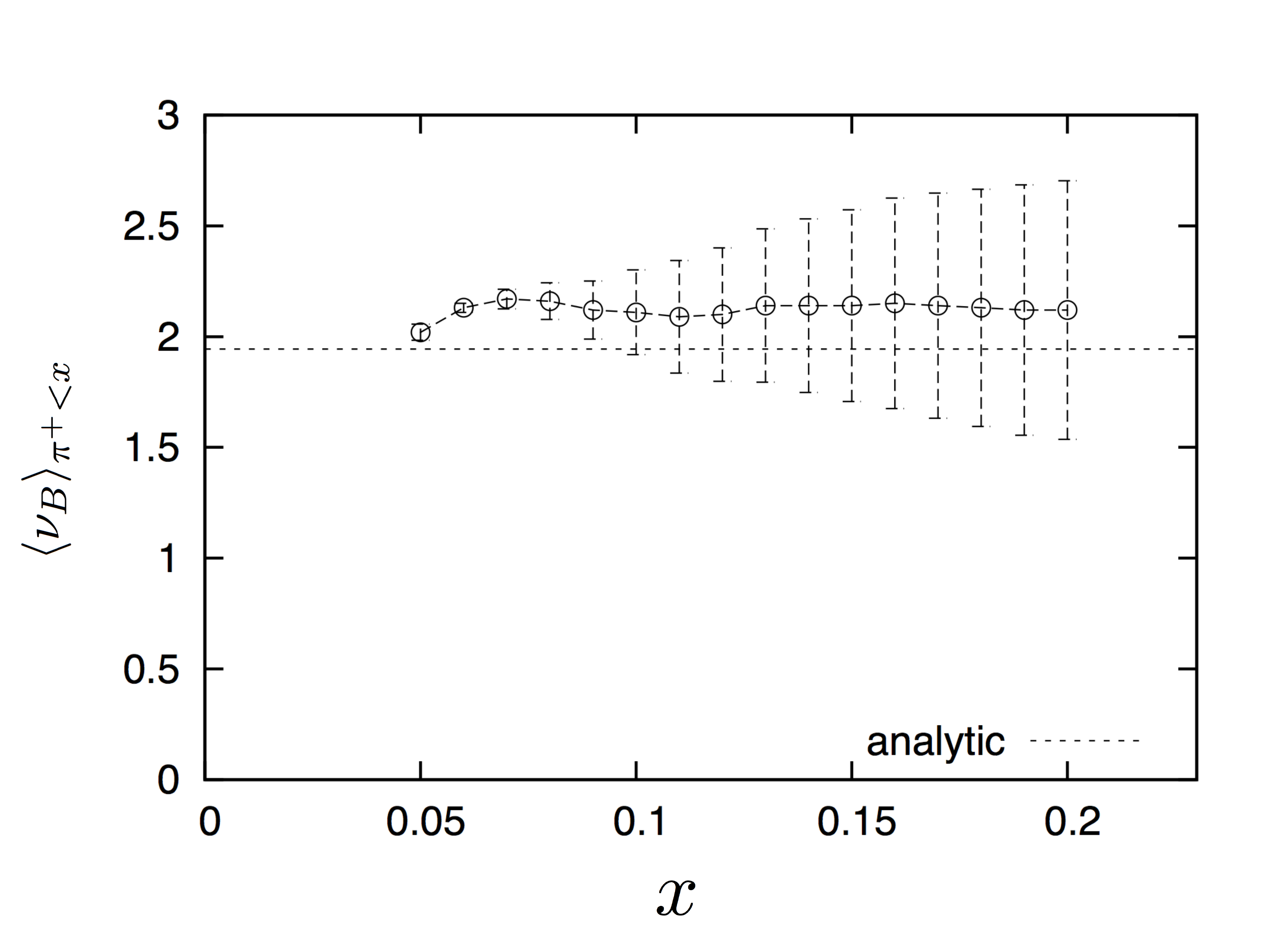}}}
   \end{center}
   \caption{$\langle \nu_B\rangle_{\pi^+<x}$ calculated at limited range of $\pi^+$. 
   $N=8$, $m=0$, $\mu=0.7$ and $c=0.02$.   
 }
\label{fig:N8M000C070_baryon_density}
\end{figure}

Next let us consider $N=8$, $m=0$, $\mu=0.7$ and $c=0.02$. 
We take $\epsilon=0.01$, $x=0.01, 0.02, 0.03,\cdots$. 
At $x\ge 0.05$, we collected $10,000,000$ -- $13,800,000$ configurations for each bin. 
In the chiral limit, the phase fluctuation becomes severer as $N$ increases: see Fig.~\ref{fig:N8M000C070_phase} in which the average phase for $N=8$ and $N=4$ are shown. 
Still, the sign problem can be controlled by fixing $\pi^+$ to be small.  
In Fig.~\ref{fig:N8M000C070S002RelativeWeight} we show $\rho_i\cdot\langle {\rm e}^{iS_I}\rangle_i$. 
We can see the dominant contribution comes from the small-$\pi^+$ region. 
It is reasonable to omit configurations with $\pi^+\gtrsim 0.12$, and there we can evaluate the baryon density reasonably well, 
as shown in Fig.~\ref{fig:N8M000C070_baryon_density}.

\subsection{Fixing $\nu_B$}\label{sec:fix_nu}
\hspace{0.51cm}

As we have mentioned in Sec.~\ref{sec:comment_baryon_pin_dome}, at $\mu<\mu_c$, 
the baryon density $\nu_B$ could be used to pin down the correct vacuum. 
(More precisely, we fix the real part, ${\rm Re}[\nu_B]$.)
So let us see the correlation between the average phase and $\nu_B$ at $\mu=0.4$, which is below $\mu_c$, 
and at $\mu=0.7$, which is above $\mu_c$. 
\begin{figure}[htbp]
   \begin{center}\rotatebox{0}{
   \scalebox{1.5}{
     \includegraphics[height=6cm]{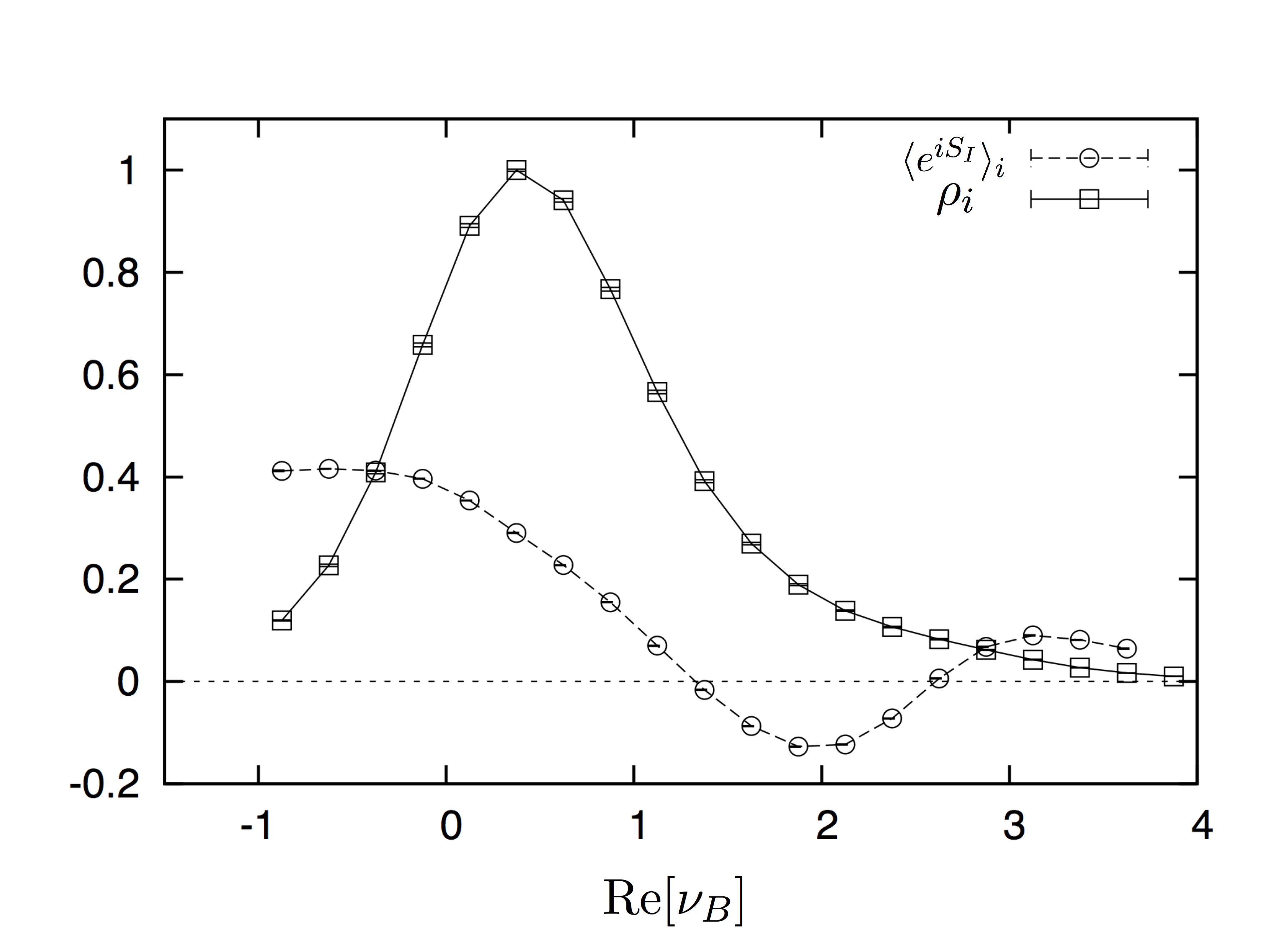}}}
   \end{center}
   \caption{Histogram of ${\rm Re}[\nu_B]$ and the average phase as a function of $\nu_B$ at $N=4$, $m=0$ and $\mu=0.4$. 
   }
\label{fig:N4M000_pin_dome_baryon_mu=040}
\end{figure} 
\begin{figure}[htbp]
   \begin{center}\rotatebox{0}{
   \scalebox{1.5}{
     \includegraphics[height=6cm]{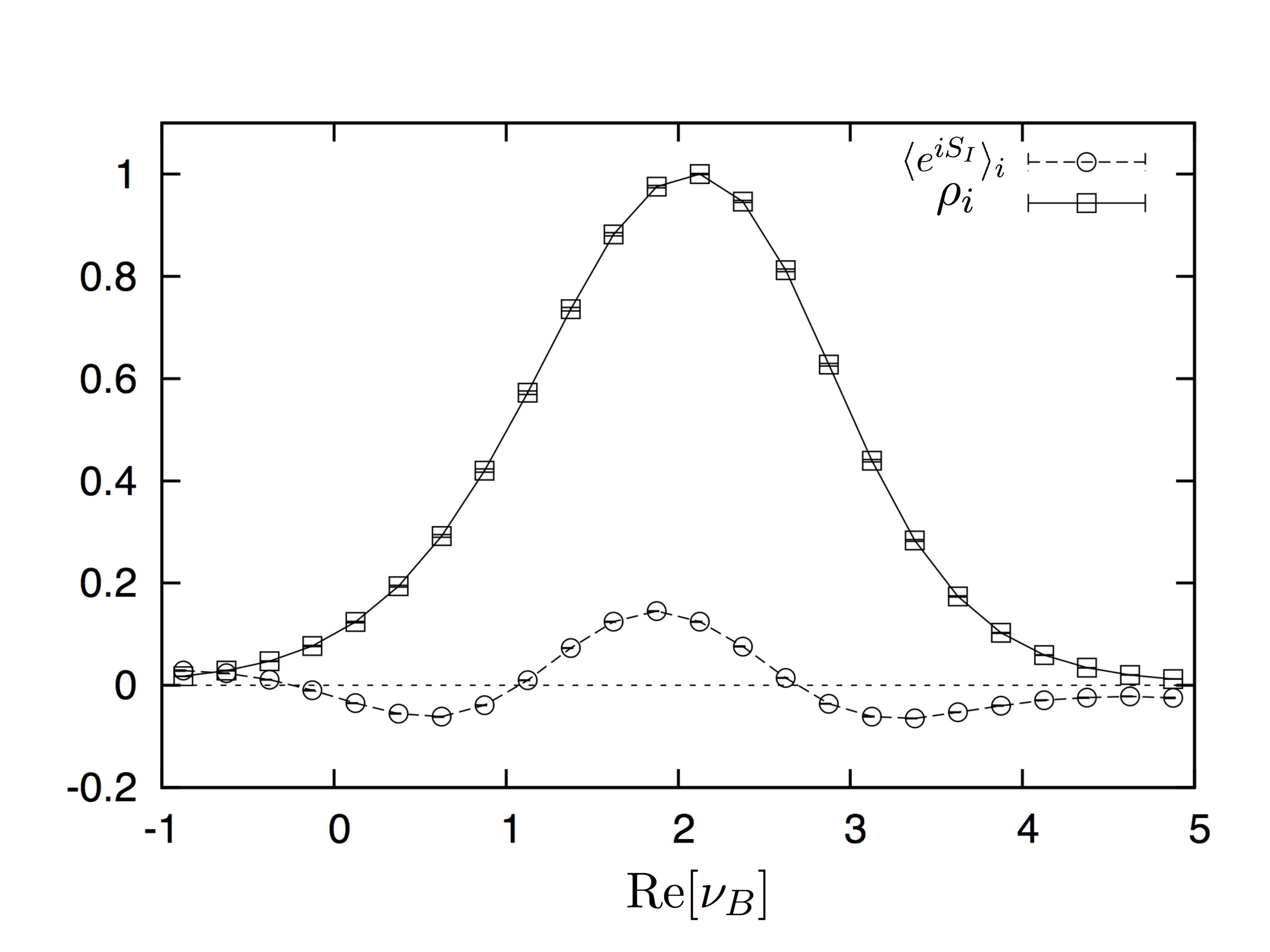}}}
   \end{center}
   \caption{Histogram of ${\rm Re}[\nu_B]$ and the average phase as a function of $\nu_B$ at  $N=4$, $m=0$ and $\mu=0.7$. 
   }
\label{fig:N4M000_pin_dome_baryon_mu=070}
\end{figure} 
In Fig.~\ref{fig:N4M000_pin_dome_baryon_mu=040}, the histogram of the real part of the baryon density ${\rm Re}[\nu_B]$, and average phase at $\mu=0.4$ are shown. 
The average phase is larger at small ${\rm Re}[\nu_B]$ region as expected. However, the average phase remain non-negligible even at large ${\rm Re}[\nu_B]$ region, 
which suggests the baryon density is not as good observable as the pion condensate,
though it could be used to make the corresponding density of states at $\mu < \mu_c$.

\begin{figure}[htbp]
   \begin{center}\rotatebox{0}{
   \scalebox{1.5}{
     \includegraphics[height=6.5cm]{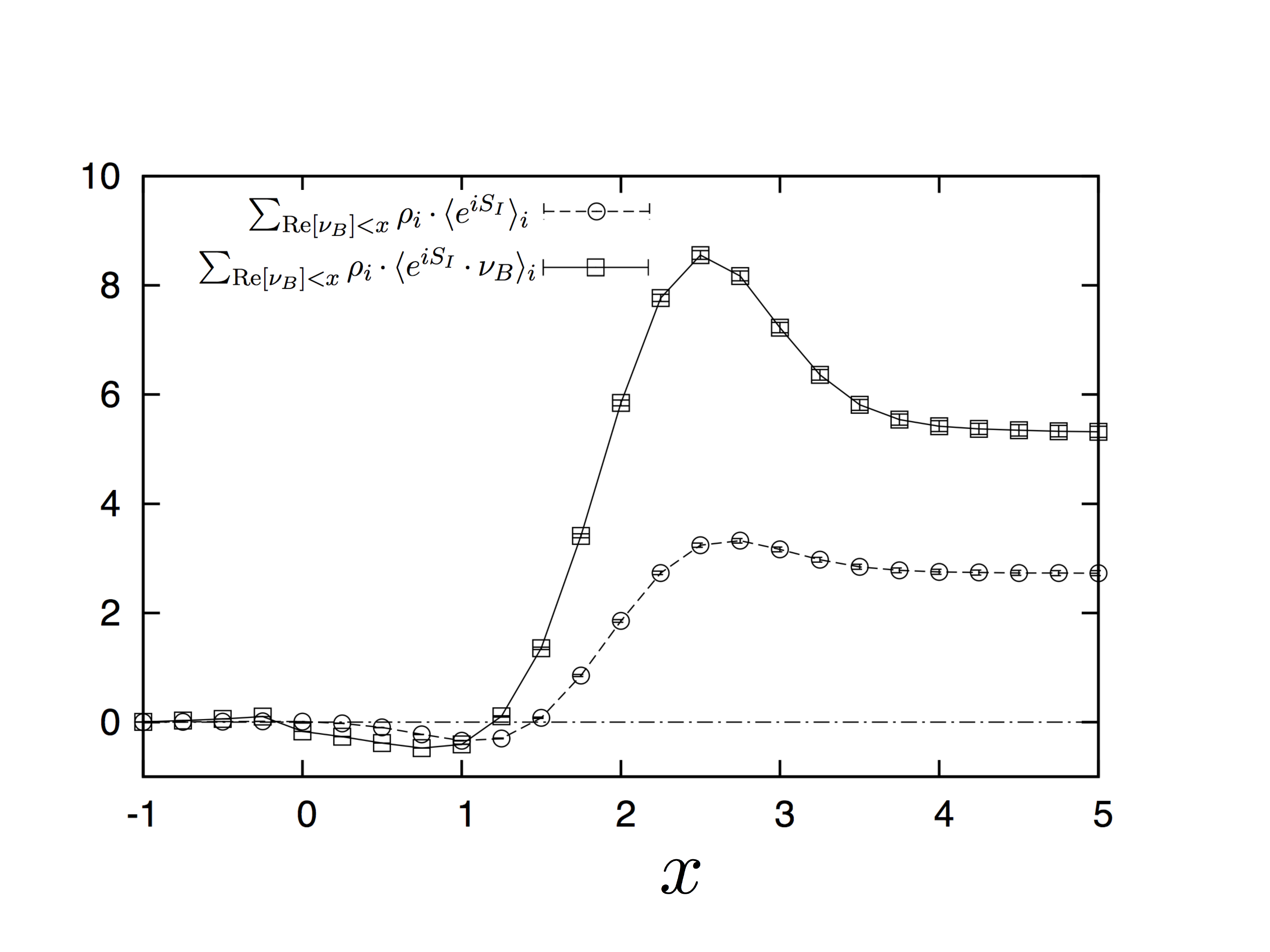}}}
   \end{center}
   \caption{ $\sum_{{\rm Re}[\nu_B]<x} \rho_i\cdot \langle {\rm e}^{iS_I}\rangle_i$ and $\sum_{{\rm Re}[\nu_B]<x} \rho_i \cdot \langle {\rm e}^{iS_I}\cdot \nu_B\rangle_i$ calculated at limited range of ${\rm Re}\nu_B$. 
   $N=4$, $m=0$, $\mu=0.7$ and $c=0.02$.  
  The peak of $\rho_i \cdot\langle {\rm e}^{iS_I}\rangle_i$ is normalized to be 1. }
\label{fig:N4M000_pin_dome_baryon_mu=070_sum_phase}
\end{figure}

\begin{figure}[htbp]
   \begin{center}\rotatebox{0}{
   \scalebox{0.8}{
     \includegraphics[height=6.5cm]{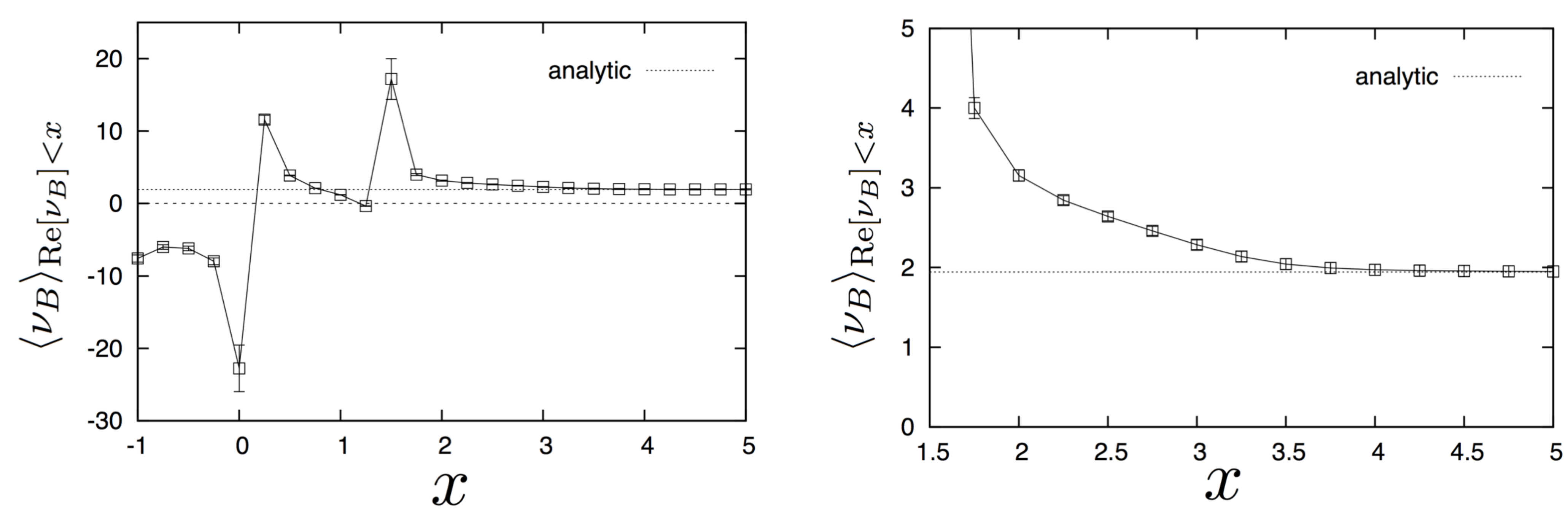}}}
   \end{center}
   \caption{$\langle\nu_B\rangle_{{\rm Re}[\nu_B] < x}$ at  $N=4$, $m=0$ and $\mu=0.7$. 
   The right panel is the zoom-up of a part of the left one. 
   }\label{fig:nuB_x}
\end{figure}

Moreover, as shown in Fig.~\ref{fig:N4M000_pin_dome_baryon_mu=070}, 
at $\mu=0.7$ the average phase oscillates around zero, a very complicated cancellation takes place and hence 
one has to study whole the configurations in order to estimate  $\langle\nu_B\rangle$ precisely.
In order to make this point clearer, we show $\sum_{{\rm Re}[\nu_B]<x} \rho_i\cdot \langle {\rm e}^{iS_I}\rangle_i$ and 
$\sum_{{\rm Re}[\nu_B]<x} \rho_i \cdot \langle {\rm e}^{iS_I}\cdot \nu_B\rangle_i$ in Fig.~\ref{fig:N4M000_pin_dome_baryon_mu=070_sum_phase}. 
We can see a large $x$-dependence at $0\lesssim x\lesssim 4$. 
As a result, the convergence of $\langle\nu_B\rangle_{{\rm Re}[\nu_B] < x}$ is very slow, as shown in Fig.~\ref{fig:nuB_x}. 
$\langle\nu_B\rangle_{{\rm Re}[\nu_B] < x}$ becomes close enough to the correct value of $\langle\nu_B\rangle$ 
only at $x\gtrsim 4$, where almost all the configurations in the phase-quenched simulation are considered. 
Therefore $\nu_B$ is not an appropriate observable to single out the correct vacuum at $\mu > \mu_c$. 

\section{Strategies for the full QCD simulations}
\label{sec:QCD}
\hspace{0.51cm}
In this section we discuss a few strategies to apply our idea to the full QCD simulations. 
\subsection{Low-$T$, large-$\mu$ region}
\hspace{0.51cm}
In the low temperature and high density region, one has to overcome the pion condensate by introducing the constraint term like \eqref{deformation_RMT}. 
However since the simulation cost is not small, one needs to choose the constraint term in a clever manner so that efficient algorithms 
e.g. the Hybrid Monte Carlo (HMC) method are applicable. 
For that purpose, we introduce the gaussian term 
\begin{eqnarray}
\Delta S 
=
\gamma\int d^4x |\pi^+(x) - a|^2 
\end{eqnarray}
again, but this time 
we do not set it to zero near $a$. Instead we take the above Gaussian form for all values of $\pi^+(x)$. 
(And, again, we introduce the source only for $\Delta S$.)
This four-fermi term can be made fermion bi-linear by introducing an auxiliary field, which allows us to apply the HMC method. 
A simple method for reconstructing the histogram of $\pi^+$ in the phase-quenched simulation with this deformation term 
can be found in \cite{Anagnostopoulos:2001yb}.

\subsection{High-$T$, small-$\mu$ region}\label{sec:low_T_high_mu}
\hspace{0.51cm}
The overlap problem is not severe at high-$T$, small-$\mu$ region. 
Still, at large volume, the sign fluctuation becomes very violent and simulation cost increases. 

The origin of the overlap problem in this region is the gas of charged pion. 
Since the pion is light, the gas of pions can easily be excited with the isospin chemical potential, 
and hence the isospin density $\nu_I$ takes non-negligible value.  
On the other hand, with the baryon chemical potential, only the gas of baryons can be excited. 
Since baryons are heavy, the baryon density $\nu_B$ must be small. 
By recalling $\nu_B$ in QCD$_B$ corresponds to $\nu_I$ in QCD$_I$  
(i.e. $\langle\nu_B\rangle_B=\langle\nu_I \cdot {\rm e}^{iS_I}\rangle_{I}/\langle {\rm e}^{iS_I}\rangle_{I}$), 
it is natural to think that the overlap problem can be suppressed by taking $\nu_I$ small. 

Given that the overlap problem is not severe compared to the low-$T$, large-$\mu$ region, 
important configurations with small $\nu_I$ would be contained  to some extent in the phase-quenched ensemble. 

Therefore, with the re-analysis already existing configurations by calculating $\nu_I$, classifying configurations in terms of the values of $\nu_I$,  and then applying our method, it would be possible to overcome the sign and overlapping problems.    
Note that one does not even have to calculate the determinant at large $\nu_I$, and hence it may reduce the cost for the reweighting, 
while increasing the accuracy.

\section{More generic reweighting method}
\label{sec:general}
\hspace{0.51cm}
In principle, one can consider more generic reweighting in which the reweighting factor is not just a phase. 
For example, one can use configurations generated with chemical potential $(\mu_1',\mu_2')$ to study $(\mu_1,\mu_2)$, 
\begin{eqnarray}
\langle\hat{O}\rangle_{\mu_1,\mu_2}
=
\frac{
\langle\hat{O}\cdot (\det(\mu_1,\mu_2)/\det(\mu'_1,\mu'_2))\rangle_{\mu'_1,\mu'_2}
}{
\langle \det(\mu_1,\mu_2)/\det(\mu'_1,\mu'_2)\rangle_{\mu'_1,\mu'_2}
}. 
\end{eqnarray}
Our method can easily be generalized to such cases. 

The fact that the pion condensate cause the overlap problem has been known for long long time. 
Therefore, in order to step into the pion condensation, reweighting from small-$\mu$ region has been performed. 
That is, one performed a simulation at $\mu_0 < \mu_c$, where $\mu_c$ is the critical value for the pion condensation, 
and tried to study $\mu>\mu_0$ by 
\begin{eqnarray}
\langle\hat{O}\rangle_{+\mu,+\mu}
=
\frac{
\langle\hat{O} (\det(+\mu,+\mu)/\det(+\mu_0,-\mu_0))\rangle_{+\mu_0,-\mu_0}
}{
\langle \det(+\mu,+\mu)/\det(+\mu_0,-\mu_0)\rangle_{+\mu_0,-\mu_0}
}. 
\label{eq:generic}
\end{eqnarray}
%
However this method does not solve the overlap problem, because the pion condensation at $\mu>\mu_c$ takes place 
even in such reweighting calculation. In fact, the pion condensation has been observed even in the quench simulation, 
in which the configurations are generated by using pure Yang-Mills action without fermions. 
Therefore, configurations generated at $\mu_0<\mu_c$ are not necessarily important ones at $\mu>\mu_c$; 
What one should actually do is to calculate the pion condensate $\pi^+$ at $(+\mu,-\mu)$ ({\it not} at $(+\mu_0,-\mu_0)$), 
classify the configurations and apply our method by using 
configurations with small $\pi^+$. 
Note again that one does not even have to calculate the determinant at large $\pi^+$, and hence it is possible to reduce the cost for the reweighting, 
while increasing the accuracy. 

In the phase reweighting method, once the pion condensate is set zero the sign problem is $1/N_c$ suppressed. 
For generic reweightings given in eq.~(\ref{eq:generic}), such a nice property does not exist, and hence more violent cancellation is expected. 
It should become severer as one goes deep inside the pion condensation. 
Still, however, it would be useful to study the chiral and deconfinement 
transitions by stepping a little bit inside. This can be a very important application practically, since the QCD critical point 
might be there.

\section{Conclusion and future directions}
\label{sec:conclusion}
\hspace{0.51cm}
 In this paper, we have improved the density of state method which can make the sign problem milder.  
 As a by-product, the problem of the zero-mode in the finite-density QCD, which is associated with the pion condensation in the phase quenched theory, can be avoided. 
We demonstrated our idea thoroughly by using numerically cheaper and analytically solvable toy model, the chiral random matrix theory.  

There are variety of directions for future studies. For the finite-density QCD, 
it is important to study high-$T$ low-$\mu$ region thoroughly. In addition to the confirmation of the effectiveness of the method, 
it would provide us with better numerical understanding about the nature of the QCD thermal transition.  
For this, we can use the simplified method described in Sec.~\ref{sec:low_T_high_mu} which does not require generation of new configurations. 
It is also interesting to go a little bit into the pion condensation region in order to search the QCD critical point, by using the method described in Sec.~\ref{sec:general}. 
Again, we do not need new configurations; re-analysis of existing configurations can provide us with better results. 
Needless to say, the study of the low-temperature high-density region sketched in Sec.~\ref{sec:low_T_high_mu} is the most interesting thing to do. 
Although the remaining sign problem would become severe at large volume, interesting phenomena would be seen 
already at small volume. Note that even the phase quench simulation can work up to the $1/N_c$-correction, 
once the pion condensation is erased \cite{Cherman:2010jj,Hanada:2011ju}.  

Our method is quite general. We hope to report other applications, such as models in condensed matter physics and supersymmetric gauge theory, in near future.

\section*{Acknowledgment} 
\hspace{0.51cm}
The authors would like to thank M.~Buchoff, P.~de Forcrand, S.~Ejiri, K.~Nagata, 
J.~Nishimura S.~Takeda and N.~Yamamoto for stimulating discussions and comments.
This work is supported in part by the Grant-in-Aid of the Japanese Ministry of Education, Sciences and Technology, Sports and Culture (MEXT) for Scientific Research 
(Nos. 23654092, 24340054 and 25287046) and by MEXT Strategic Program for Innovative Research (SPIRE) Field 5 and Joint Institute for Computational Fundamental Science (JICFuS).
This work was supported in part by the National Science Foundation under Grant No. PHYS-1066293 and the hospitality of the Aspen Center for Physics.

\appendix

\end{document}